\documentstyle[eqsecnum,manuscript,aps,psfig]{revtex}
\begin{document}

\newcommand{\be}{\begin{equation}}
\newcommand{\ee}{\end{equation}}
\newcommand{\bea}{\begin{eqnarray}}
\newcommand{\eea}{\end{eqnarray}}

\newcommand{\0}{\bar{0}}
\newcommand{\1}{\bar{1}}
\newcommand{\2}{\bar{2}}
\newcommand{\3}{\bar{3}}


\title{Is There Chaos in Low-Energy String Cosmology?}
\author{John D. Barrow \footnote{E-mail:J.D.Barrow@sussex.ac.uk}\\
       {\it Astronomy Centre, University of Sussex, Falmer, Brighton BN1 9QJ, U.K.,}\\
        Mariusz P. D\c{a}browski \footnote{E-mail:mpd@star.cpes.susx.ac.uk}
\\
         {\it Astronomy Centre, University of Sussex, Falmer, Brighton BN1 9QJ, U.K.,\\
Institute of Physics, University of Szczecin, Wielkopolska 15, 
          70-451 Szczecin, Poland}}
\date{\today}

\maketitle

\begin{abstract}

Bianchi type IX, 'Mixmaster' universes are investigated in
low-energy-effective-action string cosmology. We show that, unlike in general relativity,  there is no chaos in these string cosmologies for the case of the tree-level action. The characteristic Mixmaster evolution through a series of Kasner epochs is studied in detail. In the Einstein frame an infinite sequence of chaotic oscillations of the scale factors on approach to the initial singularity is impossible, as it was in general relativistic Mixmaster universes in the presence of a massless scalar field. A finite sequence of oscillations of the scale factors described by approximate Kasner metrics is possible, but it always ceases when all expansion rates become positive. In the string frame the evolution through Kasner epochs changes to a new form which reflects the duality symmetry of the theory. Again, we show that chaotic oscillations must end after a finite time. The need for duality symmetry appears to be incompatible with the presence of chaotic behaviour as $t \to 0$. We also obtain our results using the Hamiltonian qualitative cosmological picture for Mixmaster models. We also prove that a time-independent pseudoscalar axion field $h$ is not admitted by the Bianchi IX geometry.

\end{abstract}


PACS number(s): 98.80.Hw, 04.50.+h, 11.25.Mj, 98.80.Cq

\newpage

\section{Introduction}

\setcounter{equation}{0}

String cosmology has attracted a lot of interest recently, especially 
in the
context of duality symmetry, which is a striking feature of the underlying string
theory. The motion of the simplest bosonic string in background fields 
is
governed by the nonlinear sigma-model action \cite{fradcall}. 
Cosmological
solutions (to lowest order in $\alpha ^{\prime }$ - the inverse string
tension) have been considered in detail in many papers with special 
interest
in the possibility of inflation, the behaviour of inhomogeneities, and 
the
relation between the so called pre- and post-Big-Bang phases of 
evolution. 
\cite{strmod}. The bosonic string spectrum of particles contains the 
graviton, dilaton and axion (antisymmetric tensor field). The dilaton 
can
always be accommodated within homogeneous geometries, but this is not 
the case for the axion. It has been shown that the antisymmetric tensor-field
potential does not necessarily have to be time-independent and this 
can prevent 
the isotropisation of homogeneous models at late times 
\cite{ed}.
Moreover, the time-dependent antisymmetric tensor potential cannot be
accommodated in all  homogeneous anisotropic universes of Bianchi or
Kantowski-Sachs types \cite{homog}. In this paper we concentrate on 
Bianchi
IX (BIX) axion-dilaton cosmology with homogeneous ansatz as given 
originally
in \cite{bk}. This is equivalent to the inclusion of a time-dependent
pseudoscalar axion field, $h=h(t)$. The dilaton field is always 
homogeneous, $\phi =\phi(t)$.

It is well known that in the vacuum BIX homogeneous cosmology one 
approaches
the initial singularity chaotically \cite{bkl}. An 
infinite
number of oscillations of the orthogonal scale factors occurs in general on any finite interval of 
proper
time including the singularity at $t=0.$ These oscillations are created by the 
3-curvature
anisotropy of the spacetime and are intrinsically general relativistic 
in
origin. Physically, the propagation of homogeneous gravitational waves alters the
curvature of spacetime along the direction of propagation, so that their  non-linear back-reaction on the 
curvature
reverses the direction of propagation. 

In the presence of a massless 
scalar
field (or, alternatively, 'stiff matter', with pressure, $p,$ equal to
density, $\rho $) the situation changes. Only a finite number of 
spacetime
oscillations can occur before the evolution is changed into a state in 
which
all directions shrink monotonically to zero as the curvature 
singularity is
reached and the oscillatory behaviour ceases \cite{belkhal}. In this paper 
we want to investigate the oscillatory approach to singularity
in BIX low-energy-effective-action homogeneous string cosmologies with both 
axion
and dilaton fields.

This investigation complements other recent studies of the behaviour 
of
string cosmologies at early times. The investigation in \cite{veldom} 
shows
that there is an open neighbourhood of initial data space for string
cosmologies which display velocity-dominated behaviour. That is, on 
approach
to the singularity the spatial gradients become negligible compared to 
time
derivatives, velocities are non-relativistic, and curvature 
anisotropies are
negligible. These solutions resemble inhomogeneously-varying Kasner
metrics, but the approximation only hold on scales larger than the 
particle
horizon. Barrow and Kunze \cite{barkun} found a large family of exact inhomogeneous 
string
cosmologies which are velocity-dominated. They describe the evolution 
of the
dilaton and axion fields on all scales, showing how these fields 
oscillate
once inhomogeneities in their distribution enter the horizon. 
These 
studies show that the general behaviour of the low-energy-string cosmology equations are 
considerably simpler than those of general relativity. The string 
gravity
sector behaves in a quasi-Newtonian fashion in the vicinity of
cosmological singularities. In general relativity the 
velocity-dominated,
quasi-Kasner, behaviour is disrupted by the chaotic oscillations of 
the
three-curvature anisotropy on approach to the singularity whenever the 
matter
content is a massless scalar field or perfect fluid satisfying  $p<\rho .$ It is therefore 
important
to determine what happens in the presence of three-curvature 
anisotropies in
string cosmology. It is possible that there exists another stable
neighbourhood of evolution on approach to the initial singularity 
which is
chaotic rather than of quasi-Kasner form. We will investigate 
this
situation by studying the Bianchi type IX string cosmologies in order 
to
discover whether chaotic behaviour is possible or whether it is 
destroyed by
the presence of the dilaton and axion fields.

The paper is organized as follows. In Section II we derive the string
low-energy-effective-action  
equations for
BIX geometry. We present the equations in both the string and the 
Einstein
frames and we use orthonormal frames in the main body of the paper. In
Section III we discuss the possible routes to chaos near  
a curvature 
singularity in both the Einstein and the string frames with reference 
to the
calculations given in Ref.\cite{bkl}. In Section IV we discuss the 
relation
between the oscillating Mixmaster behaviour of the scale factors and the duality
symmetry of the string equations. In Section V we use the
Hamiltonian approach to represent the evolution of the BIX string  
cosmology
as the motion of a 'universe point' inside a curvature potential. This
picture provides a simple picture of the conditions needed for chaos 
to
occur, and for it to be absent, in the Einstein, string and so-called 
axion
frames. In Appendix A we give the relation between Ricci tensors for 
BIX
model in orthonormal and holonomic (coordinate) frames and  
show why it is impossible to include a time-independent pseudoscalar 
axion
field $h$ in the non-axisymmetric and axisymmetric BIX models.

\section{Low-energy-effective-action equations for Bianchi IX universes}

\setcounter{equation}{0}

The low-energy-effective-action field equations in the string frame 
are
given by \cite{ed,bk} \footnote{%
Following standard notation we use Greek indices here to write down 
the
field equations (2.1)-(2.8). However, beginning with the formula (2.9) 
we
replace these Greek indices by Latin ones $i,j,k,l=0,1,2,3$ because we 
are
using an orthonormal basis. Greek indices $\alpha ,\beta ,\mu ,\nu 
=\bar 0,%
\bar 1,\bar 2,\bar 3$ are coordinate basis indices and will be used in
Appendix A} 
\begin{eqnarray}
R_\mu ^\nu +\nabla _\mu \nabla ^\nu \phi -\frac 14H_{\mu \alpha \beta
}H^{\nu \alpha \beta } &=&0, \\
R-\nabla _\mu \phi \nabla ^\mu \phi +2\nabla _\mu \nabla ^\mu \phi 
-\frac 1{%
12}H_{\mu \nu \beta }H^{\mu \nu \beta } &=&0, \\
\nabla _\mu \left( e^{-\phi }H^{\mu \nu \alpha }\right)  &=&\partial 
_\mu
\left( e^{-\phi }\sqrt{-g}H^{\mu \nu \alpha }\right) =0,
\end{eqnarray}
where $\phi $ is the dilaton field, $H_{\mu \nu \beta }=6\partial 
_{[\mu
}B_{\nu \beta ]}$ is the field strength of the antisymmetric tensor 
$B_{\mu
\nu }=-B_{\nu \mu }$. In the Einstein frame we have \cite{ed} 
\begin{eqnarray}
\tilde R_\mu ^\nu -\frac 12\tilde g_\mu ^\nu \tilde R &=&\kappa 
^2\left( 
\tilde T_\mu ^{\nu (\phi )}+\tilde T_\mu ^{\nu (H)}\right) , \\
\tilde \nabla _\mu \tilde \nabla ^\mu \phi +\frac 16e^{-2\phi }\tilde 
H^2
&=&0, \\
\tilde \nabla _\mu \left( e^{-2\phi }\tilde H^{\mu \nu \alpha }\right) 
 &=&%
\tilde \partial _\mu \left( e^{-2\phi }\sqrt{-\tilde g}\tilde H^{\mu 
\nu
\alpha }\right) =0,
\end{eqnarray}
and ($\kappa^2 = 8\pi G, c = 1$)
\begin{eqnarray}
\kappa ^2\tilde T_\mu ^{\nu (\phi )} &=&\frac 12\left( \tilde g_\mu 
^\lambda 
\tilde g^{\nu \sigma }-\frac 12\tilde g_\mu ^\nu \tilde g^{\lambda 
\sigma
}\right) \tilde \nabla _\lambda \phi \tilde \nabla _\sigma \phi , \\
\kappa ^2\tilde T_\mu ^{\nu (H)} &=&\frac 1{12}e^{-2\phi }\left( 
3\tilde H%
_{\mu \alpha \beta }\tilde H^{\nu \alpha \beta }-\frac 12\tilde g_\mu 
^\nu 
\tilde H^2\right) .
\end{eqnarray}
Covariant derivatives are formed with respect to the Bianchi type IX  
metric
which, in the string frame, reads as \cite{j1} 
\begin{equation}
ds^2=dt^2-a^2(t)(\sigma ^1)^2-b^2(t)(\sigma ^2)^2-c^2(t)(\sigma ^3)^2,
\end{equation}
where the orthonormal forms $\sigma ^1,\sigma ^2,\sigma ^3$ are given 
by ($%
i,j,k,l=0,1,2,3$ are orthonormal basis indices) 
\begin{eqnarray}
\sigma ^1 &=&\cos {\psi }d\theta +\sin {\psi }\sin {\theta }d\varphi , 
\\
\sigma ^2 &=&\sin {\psi }d\theta -\cos {\psi }\sin {\theta }d\varphi , 
\\
\sigma ^3 &=&d\psi +\cos {\theta }d\varphi ,
\end{eqnarray}
and the angular coordinates $\psi ,\theta ,\varphi $ span the
following ranges, 
\begin{equation}
0\leq \psi \leq 4\pi ,\hspace{0.5cm}0\leq \theta \leq \pi 
,\hspace{0.5cm}%
0\leq \varphi \leq 2\pi .
\end{equation}
In the Einstein frame the metric is 
\begin{equation}
d\tilde s^2=d\tilde t^2-\tilde a^2(t)(\sigma ^1)^2-\tilde 
b^2(t)(\sigma
^2)^2-\tilde c^2(t)(\sigma ^3)^2,
\end{equation}
and 
\begin{eqnarray}
d\tilde t &=&e^{-\phi /2}dt, \\
\tilde a_i &=&e^{-\phi /2}a_i,
\end{eqnarray}
where $\tilde a_i=\{\tilde a,\tilde b,\tilde c,\}$ and 
$a_i=\{a,b,c\}$.

The nonzero Ricci tensor components in the orthonormal frame $\sigma 
^i$ are
(an overdot means a derivative with respect to the synchronous 
coordinate
time $t$) \cite{bkl} 
\begin{eqnarray}
-R_0^0 &=&\frac{\ddot a}a+\frac{\ddot b}b+\frac{\ddot c}c, \\
-R_1^1 &=&\frac{\ddot a}a+\frac{\dot a}a\frac{\dot b}b+\frac{\dot 
a}a\frac{%
\dot c}c-\frac 1{2a^2b^2c^2}\left[ \left( b^2-c^2\right) ^2-a^4\right] 
, \\
-R_2^2 &=&\frac{\ddot b}b+\frac{\dot a}a\frac{\dot b}b+\frac{\dot 
b}b\frac{%
\dot c}c-\frac 1{2a^2b^2c^2}\left[ \left( a^2-c^2\right) ^2-b^4\right] 
, \\
-R_3^3 &=&\frac{\ddot c}c+\frac{\dot a}a\frac{\dot c}c+\frac{\dot 
b}b\frac{%
\dot c}c-\frac 1{2a^2b^2c^2}\left[ \left( a^2-b^2\right) ^2-c^4\right] 
,
\end{eqnarray}
and the Ricci scalar reads 
\begin{eqnarray}
-R &=&2\frac{\ddot a}a+2\frac{\ddot b}b+2\frac{\ddot c}c+2\frac{\dot 
a}a%
\frac{\dot b}b+2\frac{\dot a}a\frac{\dot c}c+2\frac{\dot b}b\frac{\dot 
c}c 
\nonumber \\
&&-\frac 1{2a^2b^2c^2}\left[ 
a^4+b^4+c^4-2a^2b^2-2b^2c^2-2a^2c^2\right] .
\end{eqnarray}
Since the model under consideration is homogeneous the dilaton field 
can
only depend on time and so we have (cf. Appendix A) 
\begin{eqnarray}
\nabla _0\nabla ^0\phi  &=&\ddot \phi , \\
\nabla _i\nabla ^i\phi  &=&\frac{\dot a_i}{a_i}\dot \phi , \\
\nabla _0\phi \nabla ^0\phi  &=&\dot \phi ^2.
\end{eqnarray}

As for the axion field, we can use the following homogeneous ansatz 
\cite
{ed,bk} 
\begin{equation}
H=\frac A{abc}e^1\wedge e^2\wedge e^3=A\sigma ^1\wedge \sigma ^2\wedge
\sigma ^3,
\end{equation}
where $A$ is constant and 
\begin{equation}
e^i=a_i\sigma ^i.
\end{equation}
Since 
\begin{equation}
(H_i^j)^2\equiv H_{ikl}H^{jkl},
\end{equation}
and 
\begin{equation}
H^2\equiv H_{ikl}H^{ikl},
\end{equation}
we have 
\begin{eqnarray}
(H_0^0)^2 &=&0, \\
(H_1^1)^2 &=&(H_2^2)^2=(H_3^3)^2=-2\frac{A^2}{a^2b^2c^2}, \\
H^2 &=&-6\frac{A^2}{a^2b^2c^2},
\end{eqnarray}

The special case of axial symmetry $a=b$ gives the orthonormal forms
(2.10)-(2.12) as below \cite{brill} 
\begin{eqnarray}
\sigma ^1 &=&d\theta , \\
\sigma ^2 &=&\sin {\theta }d\varphi , \\
\sigma ^3 &=&d\psi +\cos {\theta }d\varphi ,
\end{eqnarray}
One can easily check (cf. Appendix A) that the homogeneous ansatz 
(2.25) is
also appropriate for the axisymmetric case.

\section{Time-dependent pseudoscalar axion field and the oscillatory
approach to a singularity}

\setcounter{equation}{0}

The ansatz (2.25) is, in fact, equivalent to the inclusion of a
time-dependent pseudoscalar axion field $h=h(t)$ \cite{ed} (cf. 
Appendix A).
We are going to investigate the possible emergence of chaos in such 
models.
Using (2.17)-(2.24) and (2.29)-(2.31) the field equations (2.1) in the
string frame read ($^{\cdot }=d/dt$) 
\begin{eqnarray}
\frac{\ddot a}a+\frac{\ddot b}b+\frac{\ddot c}c-\ddot \phi  &=&0, \\
\frac{\ddot a}a+\frac{\dot a}a\frac{\dot b}b+\frac{\dot a}a\frac{\dot 
c}c-%
\frac{\dot a}a\dot \phi -\frac 1{2a^2b^2c^2}\left[ \left( 
b^2-c^2\right)
^2-a^4\right] -\frac 12\frac{A^2}{a^2b^2c^2} &=&0, \\
\frac{\ddot b}b+\frac{\dot a}a\frac{\dot b}b+\frac{\dot b}b\frac{\dot 
c}c-%
\frac{\dot b}b\dot \phi -\frac 1{2a^2b^2c^2}\left[ \left( 
a^2-c^2\right)
^2-b^4\right] -\frac 12\frac{A^2}{a^2b^2c^2} &=&0, \\
\frac{\ddot c}c+\frac{\dot a}a\frac{\dot c}c+\frac{\dot b}b\frac{\dot 
c}c-%
\frac{\dot c}c\dot \phi -\frac 1{2a^2b^2c^2}\left[ \left( 
a^2-b^2\right)
^2-c^4\right] -\frac 12\frac{A^2}{a^2b^2c^2} &=&0.
\end{eqnarray}
Eq. (2.2), with Eq. (3.1), reads \cite{homog,bk} 
\begin{eqnarray}
-2\left( \frac{\dot a}a\frac{\dot b}b+\frac{\dot a}a\frac{\dot 
c}c+\frac{%
\dot b}b\frac{\dot c}c\right) +2\dot \phi \left( \frac{\dot 
a}a+\frac{\dot b}%
b+\frac{\dot c}c\right)  &-&\dot \phi ^2  \nonumber \\
+\frac 1{2a^2b^2c^2}\left[ a^4+b^4+c^4-2a^2b^2-2a^2c^2-2b^2c^2\right] 
 &+&%
\frac 12\frac{A^2}{a^2b^2c^2}=0.
\end{eqnarray}
Using (3.5), together with the sum of (3.2)-(3.4), we have the dilaton
equation of motion 
\begin{equation}
\ddot \phi -(\dot \phi )^2+\left( \frac{\dot a}a+\frac{\dot 
b}b+\frac{\dot c}%
c\right) \dot \phi -\frac{A^2}{a^2b^2c^2}=0.
\end{equation}
In the Einstein frame, the field equations (2.4) become \cite{ed} 
($^{\prime
}=d/d\tilde t$) 
\begin{eqnarray}
-\frac{\tilde a^{\prime \prime }}{\tilde a}-\frac{\tilde b^{\prime 
\prime }}{%
\tilde b}-\frac{\tilde c^{\prime \prime }}{\tilde c} &=&\frac 12\phi
^{\prime 2}+\frac 12A^2\frac{e^{-2\phi }}{\tilde a^2\tilde b^2\tilde 
c^2}, \\
\frac{\tilde a^{\prime \prime }}{\tilde a}+\frac{\tilde a^{\prime 
}}{\tilde a%
}\frac{\tilde b^{\prime }}{\tilde b}+\frac{\tilde a^{\prime }}{\tilde 
a}%
\frac{\tilde c^{\prime }}{\tilde c} &=&\frac 1{2\tilde a^2\tilde 
b^2\tilde c%
^2}\left[ \left( \tilde b^2-\tilde c^2\right) ^2-\tilde a^4\right] , 
\\
\frac{\tilde b^{\prime \prime }}{\tilde b}+\frac{\tilde a^{\prime 
}}{\tilde a%
}\frac{\tilde b^{\prime }}{\tilde b}+\frac{\tilde b^{\prime }}{\tilde 
c}%
\frac{\tilde c^{\prime }}{\tilde c} &=&\frac 1{2\tilde a^2\tilde 
b^2\tilde c%
^2}\left[ \left( \tilde a^2-\tilde c^2\right) ^2-\tilde b^4\right] , 
\\
\frac{\tilde c^{\prime \prime }}{\tilde c}+\frac{\tilde c^{\prime 
}}{\tilde c%
}\frac{\tilde b^{\prime }}{\tilde b}+\frac{\tilde a^{\prime }}{\tilde 
a}%
\frac{\tilde c^{\prime }}{\tilde c} &=&\frac 1{2\tilde a^2\tilde 
b^2\tilde c%
^2}\left[ \left( \tilde a^2-\tilde b^2\right) ^2-\tilde c^4\right] ,
\end{eqnarray}
and (2.5) now becomes 
\begin{equation}
\phi ^{\prime \prime }+\phi ^{\prime }\left( \frac{\tilde a^{\prime 
}}{%
\tilde a}+\frac{\tilde b^{\prime }}{\tilde b}+\frac{\tilde c^{\prime 
}}{%
\tilde c}\right) =A^2\frac{e^{-2\phi }}{\tilde a^2\tilde b^2\tilde 
c^2}.
\end{equation}

A new time coordinate is introduced to simplify the field equations by
defining \cite{ed,bk} 
\begin{equation}
d\eta =\frac{e^\phi }{abc}dt=\frac 1{\tilde a\tilde b\tilde c}d\tilde 
t.
\end{equation}
First, notice that the string-frame equation (3.6) then simplifies to 
\begin{equation}
\phi _{,\eta \eta }-A^2e^{-2\phi }=0,
\end{equation}
where $(,_{\eta} = d/d\eta)$.
The Einstein-frame Eq.(3.11), with the time coordinate $\tilde t,$ 
gives the
same result, (3.13). The solution of (3.13) is \cite{bk} 
\begin{equation}
e^\phi =\cosh {\Lambda M\eta }+\sqrt{1-\frac{A^2}{\Lambda ^2M^2}}\sinh 
{%
\Lambda M\eta },
\end{equation}
with $M,\Lambda $ constant ($\Lambda ^2M^2>A^2$). A useful relation, 
implied
by (3.14) is 
\begin{equation}
\phi _{,\eta \eta }+\phi _{,\eta }^2=\Lambda ^2M^2.
\end{equation}
Let us introduce new forms for the scale factors 
\begin{equation}
a=e^\alpha \hspace{2.0cm}b=e^\beta \hspace{2.0cm}c=e^\gamma ,
\end{equation}
and 
\begin{equation}
\tilde a=e^{\tilde \alpha }\hspace{2.0cm}\tilde b=e^{\tilde \beta }%
\hspace{2.0cm}\tilde c=e^{\tilde \gamma },
\end{equation}
so, from (2.16) 
\begin{equation}
\tilde \alpha =\alpha -\phi /2\hspace{2.0cm}\tilde \beta =\beta -\phi 
/2%
\hspace{2.0cm}\tilde \gamma =\gamma -\phi /2,
\end{equation}
and $dt=\exp {(\phi /2)}d\tilde t$ in (3.12). The field equations
(3.1)-(3.4) in the string frame take the form 
\begin{eqnarray}
\left( \alpha +\beta +\gamma \right) _{, \eta \eta }-M^2 
&=&2\left(
\alpha _{, \eta }\beta _{, \eta }+\alpha _{, \eta 
}\gamma
_{, \eta }+\beta _{, \eta }\gamma _{, \eta }\right) 
-2\left(
\alpha _{, \eta }+\beta _{\prime \eta }+\gamma _{, \eta 
}\right)
\phi _{\prime \eta }, \\
2e^{2\phi }\alpha _{, \eta \eta } &=&\left( b^2-c^2\right) 
^2-a^4+A^2,
\\
2e^{2\phi }\beta _{, \eta \eta } &=&\left( a^2-c^2\right) 
^2-b^4+A^2, \\
2e^{2\phi }\gamma _{, \eta \eta } &=&\left( a^2-b^2\right) 
^2-c^4+A^2.
\end{eqnarray}
The equations (3.20)-(3.22), using (3.13), can be rewritten to give 
\begin{eqnarray}
\left( -\phi +2\alpha \right) _{, \eta \eta } &=&\left[ \left(
b^2-c^2\right) ^2-a^4\right] e^{-2\phi }, \\
\left( -\phi +2\beta \right) _{, \eta \eta } &=&\left[ \left(
a^2-c^2\right) ^2-b^4\right] e^{-2\phi }, \\
\left( -\phi +2\gamma \right) _{, \eta \eta } &=&\left[ \left(
a^2-b^2\right) ^2-c^4\right] e^{-2\phi }.
\end{eqnarray}
Notice that there is no explicit dependence of the axion, $A$, in these 
equations (3.23)-(3.25).

On the other hand, using (3.12), Eqs.(3.7)-(3.10) in the Einstein 
frame
reduce to 
\begin{eqnarray}
2\left( \tilde \alpha _{, \eta }\tilde \beta _{, \eta 
}+\tilde 
\alpha _{, \eta }\tilde \gamma _{, \eta }+\tilde \beta 
_{, 
\eta }\tilde \gamma _{, \eta }\right)  &=&\left( \tilde \alpha 
+\tilde 
\beta +\tilde \gamma \right) _{, \eta \eta }+\frac 12M^2, \\
2\tilde \alpha _{, \eta \eta } &=&\left( \tilde b^2-\tilde 
c^2\right)
^2-\tilde a^4, \\
2\tilde \beta _{, \eta \eta } &=&\left( \tilde a^2-\tilde 
c^2\right) ^2-%
\tilde b^4, \\
2\tilde \gamma _{, \eta \eta } &=&\left( \tilde a^2-\tilde 
b^2\right)
^2-\tilde c^4.
\end{eqnarray}
These equations (except for the constant $M$ which appears in (3.26)) have 
exactly the
same form as the standard BIX equations of general relativity 
\cite{bkl}. Equations (3.26)-(3.29) can be explicitly transformed into
(3.23)-(3.25) by using (3.16)-(3.18).

In order to discuss possible emergence of chaos we  consider suitable 
initial
conditions expressed in terms of the Kasner parameters.

\subsection{Einstein Frame}

The Kasner solutions are obtained as approximate solutions of the 
equations
(3.7)-(3.11) when the right-hand sides (describing the curvature anisotropies) are neglected. In the Einstein 
frame,
in terms of $\tilde t$-time, they are 
\begin{eqnarray}
\tilde a &=&\tilde a_0\tilde t^{\tilde p_1},  \nonumber \\
\tilde b &=&\tilde b_0\tilde t^{\tilde p_2}, \\
\tilde c &=&\tilde c_0\tilde t^{\tilde p_3},  \nonumber
\end{eqnarray}
while ($A^2\le \tilde \Lambda ^2M^2$) 
\begin{equation}
\phi =\ln {\tilde d_0}+\frac M{\tilde \Lambda }\ln {|\tilde t|},
\end{equation}
for $A=0$, and 
\begin{equation}
\phi =\ln {\tilde d_0}+\ln {\left[ \frac 12\left( 
1+\sqrt{1-\frac{A^2}{%
\tilde \Lambda M^2}}\right) |\tilde t|^{\frac M{\tilde \Lambda 
}}+\frac 12%
\left( 1-\sqrt{1-\frac{A^2}{\tilde \Lambda M^2}}\right) |\tilde 
t|^{-\frac M{%
\tilde \Lambda }}\right] },
\end{equation}
for $A\neq 0$, where 
\begin{equation}
\tilde \Lambda =\tilde a_0\tilde b_0\tilde c_0,\hspace{2.cm}\tilde 
d_0={\rm %
const.}
\end{equation}

{}From (3.8)-(3.10), irrespective to the presence of the axion term $A$, we 
have
the following algebraic conditions on the Kasner indices, $\tilde 
p_i:$ 
\begin{equation}
\sum_{i=1}^3\tilde p_i=1,
\end{equation}
and, from (3.7), 
\begin{equation}
\sum_{i=1}^3\tilde p_i^2=1-\ \frac{M^2}{2\tilde {\Lambda ^2}},
\end{equation}
which is exactly the case considered first by Belinskii and 
Khalatnikov \cite
{belkhal} \footnote{In their notation it happens if we put $M^2/2\tilde \Lambda ^2=q^2$. Also, 
they define the Lagrangian of the scalar field without a factor of 
$1/2$
(i.e., $L_{\phi} = (\nabla \phi)^2$) while we follow standard 
notation.}.
So, in the Einstein frame their argument follows.

In terms of $\eta$-time from (3.27)-(3.29) 
\begin{eqnarray}
\tilde{\alpha} & = & \tilde{\Lambda}\tilde{p}_1 \eta + \tilde{r}_1 , 
\nonumber \\
\tilde{\beta} & = & \tilde{\Lambda}\tilde{p}_2 \eta + \tilde{r}_2 , \\
\tilde{\gamma} & = & \tilde{\Lambda}\tilde{p}_3 \eta + \tilde{r}_3 , 
\nonumber
\end{eqnarray}
and the constraint equation (3.26) bocomes \footnote{%
In \cite{bk} there is $A^2$ instead of $M^2$ while in all our 
calculations
we obtained $M^2$ both in the string frame and in the Einstein frame. 
They claim they use the so-called `sigma frame', which, however, 
seems to
be just the Einstein frame \cite{linde}.} 
\begin{eqnarray}
\tilde{p}_1 \tilde{p}_2 + \tilde{p}_1 \tilde{p}_3 + \tilde{p}_2 
\tilde{p}_3
= \frac{1}{4} \frac{M^2}{\tilde{\Lambda^2}} .
\end{eqnarray}
This constraint is equivalent to the constraints 
(3.34)-(3.35). In
order to get (3.30) from (3.36) we need to use (3.12) to relate 
$\tilde{t}$%
-time with $\eta$-time, i.e., 
\begin{equation}
\eta = \tilde{\Lambda}^{-1} \ln{\tilde{t}} + {\rm const} .
\end{equation}

When we approach singularity $\eta \to -\infty $ $(\tilde t\to 0)$ we 
cannot
neglect all the terms on the right-hand side of (3.27)-(3.29) since 
one of
the Kasner indices in (3.37) can be negative. We assume $\tilde 
p_1\equiv -|%
\tilde p_1|<0$, which means that $\tilde a(\eta )\gg \tilde b(\eta 
)\gg 
\tilde c(\eta )$, so (3.27)-(3.29) are approximated by  
\begin{eqnarray}
\tilde \alpha _{, \eta \eta } &=&-\frac 12e^{4\tilde \alpha }, 
\nonumber \\
\tilde \beta _{\prime \eta \eta } &=&\frac 12e^{4\tilde \alpha }, \\
\tilde \gamma _{\prime \eta \eta } &=&\frac 12e^{4\tilde \alpha }.  
\nonumber
\end{eqnarray}

Far away from singularity the approximate axisymmetric Kasner regime 
is
fulfilled, but it is broken by the $\tilde a^2$ term when we approach
singularity, so our Kasner solutions (3.36) will be fulfilled for 
$\eta \to
\infty $ ($\tilde t\to \infty $); suitable solutions of (3.39) 
which
fulfil (3.36) are \footnote{%
This necessarily requires $\tilde{p_1} < 0$. Of course, one has the 
same
solutions for $\tilde{p_1} > 0$, but in such a case our initial 
conditions
would have to be taken at $\eta \to \infty$ rather than at $\eta \to - 
\infty$.
This does not change the physics of the problem and was assumed in 
\cite
{halp,ishihara}, but we prefer to follow the spirit of \cite{bkl}.} 
\begin{eqnarray}
\tilde \alpha (\eta ) &=&-\frac 12\ln {\left[ \frac 1{2|\tilde 
p_1|\tilde 
\Lambda }\cosh {\left( -2|\tilde p_1|\tilde \Lambda \eta \right) 
}\right] },
\nonumber \\
\tilde \beta (\eta ) &=&\frac 12\ln {\left[ \frac 1{2|\tilde 
p_1|\tilde 
\Lambda }\cosh {\left( -2|\tilde p_1|\tilde \Lambda \eta \right) 
}\right] }+%
\tilde \Lambda \left( -|\tilde p_1|+\tilde p_2\right) \eta , \\
\tilde \gamma (\eta ) &=&\frac 12\ln {\left[ \frac 1{2|\tilde 
p_1|\tilde 
\Lambda }\cosh {\left( -2|\tilde p_1|\tilde \Lambda \eta \right) 
}\right] }+%
\tilde \Lambda \left( -|\tilde p_1|+\tilde p_3\right) \eta .  
\nonumber
\end{eqnarray}

In the limit $\eta \to - \infty$ ($\tilde{t} \to 0$) from (3.40) we 
have 
\begin{eqnarray}
\tilde{\alpha}(\eta) & = & \tilde{\Lambda} \vert \tilde{p}_1 \vert 
\eta = - 
\tilde{\Lambda} \tilde{p}_1 \eta ,  \nonumber \\
\tilde{\beta}(\eta) & = & \tilde{\Lambda} \left( \tilde{p}_2 - 2 \vert 
\tilde{p}_1 \vert \right) \eta = \tilde{\Lambda} \left( \tilde{p}_2 + 
2 
\tilde{p}_1 \right) \eta , \\
\tilde{\gamma}(\eta) & = & \tilde{\Lambda} \left( \tilde{p}_2 - 2 
\vert 
\tilde{p}_1 \vert \right) \eta = \tilde{\Lambda} \left( \tilde{p}_3 + 
2 
\tilde{p}_1 \right) \eta ,  \nonumber
\end{eqnarray}
which means that one Kasner epoch (3.36) with indices $\tilde{p}_i$ is
replaced by another Kasner epoch with indices given by (3.41). By 
virtue of
(3.12) 
\begin{equation}
\eta = \tilde{\Lambda}^{\prime -1} \ln{\tilde{t}} + {\rm const.} ,
\end{equation}
and 
\begin{eqnarray}
\tilde{a} & = & \tilde{a}_0^{\prime} \tilde{t}^{\tilde{p}_1^{\prime}} 
, 
\nonumber \\
\tilde{b} & = & \tilde{b}_0^{\prime} \tilde{t}^{\tilde{p}_2^{\prime}} 
, \\
\tilde{c} & = & \tilde{c}_0^{\prime} \tilde{t}^{\tilde{p}_3^{\prime}} 
, 
\nonumber
\end{eqnarray}
where 
\begin{eqnarray}
\tilde{p}_1^{\prime} & = & - \frac{\vert \tilde{p}_1 \vert}{1 - 2 
\vert 
\tilde{p}_1 \vert} = \frac{\tilde{p}_1}{1 + 2 \tilde{p}_1} ,  
\nonumber \\
\tilde{p}_2^{\prime} & = & \frac{\tilde{p}_2 - 2 \vert \tilde{p}_1 
\vert} {1
- 2 \vert \tilde{p}_1 \vert} = \frac{\tilde{p}_2 + 2\tilde{p}_1} {1 + 
2 
\tilde{p}_1} , \\
\tilde{p}_3^{\prime} & = & \frac{\tilde{p}_3 - 2 \vert \tilde{p}_1 
\vert} {1
+ 2 \tilde{p}_1} = \frac{\tilde{p}_3 + 2\tilde{p}_1}{1 + 2 
\tilde{p}_1} , 
\nonumber
\end{eqnarray}
and 
\begin{eqnarray}
\tilde{\Lambda}^{\prime} \tilde{t} = 
\tilde{a}\tilde{b}\tilde{c}\tilde{d} , %
\hspace{2.cm} \frac{\tilde{\Lambda}^{\prime}}{\tilde{\Lambda}} = 1 + 
2\tilde{%
p}_1 ,  \nonumber \\
\sum_{i=1}^{3} \tilde{p}_i^{\prime} = 1 , \hspace{2.cm} \sum_{i=1}^{3} 
\tilde{p}_i^{\prime 2} = 1 - \frac{1}{2} \frac{M^2} 
{\tilde{\Lambda}^{\prime
2}} .
\end{eqnarray}
As we can see from (3.32), the axion does not influence these asymptotic solutions.

One can easily show that 
\begin{equation}
-\sqrt{\frac 23}\leq \frac M{\tilde \Lambda \sqrt{2}}\leq \sqrt{\frac 
23}
\end{equation}
and, after ordering the Kasner indices by $\tilde p_1\leq \tilde p_2\leq 
\tilde p_3$, we require  
\begin{eqnarray}
-1/3 &\leq &\tilde p_1\leq 1/3,  \nonumber \\
0 &\leq &\tilde p_2\leq 2/3, \\
1/3 &\leq &\tilde p_3\leq 1.  \nonumber
\end{eqnarray}
This means that, unlike the vacuum case, all the Kasner 
indices can be 
positive and, as in the analysis of \cite{belkhal}, the final
situation is that the universe inevitably reaches a monotonic stage of 
evolution in which 
all three Kasner indices are positive (see also the discussion of Section 
V)%
\footnote{%
Then, the terms of the type $e^{4\tilde{\alpha}}, e^{4\tilde{\beta}}, 
e^{4%
\tilde{\gamma}}$ decrease if $\eta \to -\infty$ ($t \to 0$), and they 
do not
allow a transition into another Kasner epoch to take place. This is 
just
monotonic evolution like in isotropic Friedman case.}. By means of 
Eqs.
(3.30) and (3.32), which lead to the same relations between the Kasner
indices (3.34)-(3.35), one can extend this conclusion into the 
axion-dilaton
cosmology. The presence of the axion field cannot change the fate of 
the
universe near to a singularity and the universe  finally reaches a monotonic stage of
evolution with all three scale factors tending to zero as 
$t\rightarrow 0$.
Thus, there is no chaos in BIX homogeneous string cosmology in the 
Einstein
frame. Finally, we note that there is an isotropic limit here, 
provided all
the Kasner indices are equal with $\tilde p_1=\tilde p_2=\tilde p_3=1/3$.

\subsection{String Frame}

For the string frame we can follow the discussion given in \cite{halp} (see
also \cite{john85,ishihara}).

In terms of $t$-time, Kasner solutions of the system (3.1)-(3.4), for $A=0$,
are given by 
\begin{eqnarray}
a &=&a_0t^{p_1},  \nonumber \\
b &=&b_0t^{p_2}, \\
c &=&c_0t^{p_3},  \nonumber \\
e^{-\phi } &=&d_0t^{p_4},  \nonumber
\end{eqnarray}
or, alternatively, the last of these conditions (3.48) reads as 
\begin{equation}
\phi (t)=-\ln {d_0}-p_4\ln |t|,\hspace{0.5cm}{\rm and}\hspace{0.5cm}p_4=-M.
\end{equation}
After putting (3.48) into (3.6) we have 
\begin{equation}
p_4=1-\sum_{i=1}^3p_i,
\end{equation}
and from (3.1) we have 
\begin{equation}
\sum_{i=1}^3p_i^2=1,
\end{equation}
so (3.49) can be rewritten to give 
\begin{equation}
\phi (t)=-\ln {d_0}+(\sum_{i=1}^3p_i-1)\ln {t}.
\end{equation}
The last condition means, for instance, that a choice $p_i=(-1/3,2/3,2/3)$
gives $\sum p_i=1$ and leads to a vanishing or constant dilaton ($M=0$),
i.e., to general relativity. (Other permutations of the three $p_i$ with the
signs changed are allowed). Thus, the difference between general relativity and string theory depends on the ''fourth Kasner index'' $p_4= - M$.

Now, from (3.23)-(3.25) (one can see that $\Lambda p_i=2\tilde \Lambda 
\tilde p_i$ here) 
\begin{eqnarray}
\alpha  &=&\phi /2+\Lambda (p_1/2)\eta +r_1/2=\phi /2+\tilde \Lambda \tilde p%
_1\eta +\tilde r_1,  \nonumber \\
\beta  &=&\phi /2+\Lambda (p_2/2)\eta +r_2/2=\phi /2+\tilde \Lambda \tilde p%
_2\eta +\tilde r_2, \\
\gamma  &=&\phi /2+\Lambda (p_3/2)\eta +r_3/2=\phi /2+\tilde \Lambda \tilde p%
_3\eta +\tilde r_3,  \nonumber
\end{eqnarray}
and from (3.19) 
\begin{equation}
p_1p_2+p_1p_3+p_2p_3=\frac{M^2}{\Lambda ^2}.
\end{equation}

The condition (3.54), in fact, can be obtained from (3.51)-(3.52). In order
to get (3.48) we need to use (3.12), i.e., 
\begin{equation}
\eta = \Lambda^{-1} \ln{t} + {\rm const.} ,\hspace{0.5cm} \Lambda =
a_0b_0c_0d_0 .
\end{equation}

{}From the conditions (3.50)-(3.51) on the $p_i$, we have necessarily that 
\cite{belkhal} 
\begin{equation}
-1-\sqrt{3}\leq M\leq -1+\sqrt{3}.
\end{equation}
However, the domain of $M$ given by (3.57) covers the whole duality-related
region since in the case of a Kasner regime the duality symmetry (cf.
Section IV) simply means that we change 
\begin{equation}
p_1\to -p_1\hspace{0.3cm},p_2\to -p_2\hspace{0.3cm},p_3\to -p_3\hspace{0.3cm}%
,M\to M-p_1-p_2-p_3=-(M+2).
\end{equation}
Having given (3.58), we can delineate the two duality-related domains of
Kasner indices as follows: 
\begin{equation}
-1-\sqrt{3}\leq M\leq -1\hspace{0.2cm},-\frac 1{\sqrt{3}}\leq p_1\leq \sqrt{%
\frac 23}\hspace{0.3cm},-\frac 23\leq p_2\leq \frac 12\sqrt{\frac 23}%
\hspace{0.3cm},-1\leq p_3\leq -\frac 12\sqrt{\frac 23};
\end{equation}
and 
\begin{equation}
-1\leq M\leq -1+\sqrt{3}\hspace{0.2cm},-\sqrt{\frac 23}\leq p_1\leq \frac 1{%
\sqrt{3}}\hspace{0.3cm},-\frac 12\sqrt{\frac 23}\leq p_2\leq \frac 23%
\hspace{0.3cm},\frac 12\sqrt{\frac 23}\leq p_3\leq 1;
\end{equation}
Of course, for $-1-\sqrt{3}\leq M\leq -1,$ the indices are ordered so that $%
p_3\leq p_2\leq p_1$; while, for $-1\leq M\leq -1+\sqrt{3},$ they are
ordered as $p_1\leq p_2\leq p_3$.

{}From (3.57)-(3.59), we can draw some interesting conclusions. First, that
the vacuum general relativity case $M=0$ (with $-1/3\leq p_1\leq 0,0\leq p_2\leq 2/3,2/3\leq
p_3\leq 1$ ) is dual to the case $M=-2$ (with $-1\leq p_1\leq -2/3,-2/3\leq
p_2\leq 0,0\leq p_3\leq 1/3$), while the case $M=-1$ is 'self-dual' in $M$
giving $M=-1$ again, although the $p_i^{\prime }s$ change. Second, the $M=-1$ plane
is the dividing plane for the duality-related range of the parameters which
are defined by

\begin{eqnarray*}
-\sqrt{\frac{2}{3}} &\leq &p_1\leq - \frac{1}{2} \sqrt{\frac{2}{3}}, \\
-\frac{1}{2} \sqrt{\frac{2}{3}} &\leq &p_2\leq \frac{1}{2} \sqrt{\frac{2}{3}}, \\
\frac{1}{2} \sqrt{\frac{2}{3}} &\leq & p_3 \leq \sqrt{\frac{2}{3}}.
\end{eqnarray*}

The isotropic Friedmann solution is given when we choose $M=-1+\sqrt{3}%
,p_1=p_2=p_3=1/\sqrt{3}$, while its dual with $M=-1-\sqrt{3},$ $%
p_1=p_2=p_3=-1/\sqrt{3}$ does not appear in the vacuum general relativity case. This shows
that one can generally have two different types of change in the Kasner
indices, and in $M$ (effectively, a fourth Kasner index ). There are
oscillations as in general relativistic vacuum or stiff fluid cases \cite
{bkl,belkhal}, but there are also duality-related exchanges of indices. We will  discuss some points related to duality that refer to the
work of ref.\cite{wainwright}.

There are eight possible permutations of the Kasner indices in which some are equal to $ - 1/\sqrt{3}$ or $ 1/\sqrt{3}$, with suitable $M$.  These are the first two quadruples $(p_1, p_2, p_3, M) = (1/\sqrt{3}, 1/\sqrt{3}, 1/\sqrt{3}, - 1 + \sqrt{3}), ( - 1/\sqrt{3}, - 1/\sqrt{3}, - 1/\sqrt{3}, - 1 - \sqrt{3})$,  plus another three pairs with  
$(- 1/\sqrt{3}, 1/\sqrt{3}, 1/\sqrt{3}, - 1 + (\sqrt{3}/3)), (1/\sqrt{3}, - 1/\sqrt{3}, - 1/\sqrt{3}, - 1 - (\sqrt{3}/3))$ and the permutations which are dual to each other. One can easily show that these last three pairs of cases are duality related and describe exactly the transitions from one Kasner epoch to another given by (3.80). Another interesting set of 'self-dual' ($M=-1$) combinations which describe LRS (Locally Rotationally Symmetric) Kasner solutions are given by $( \sqrt{2/3}, -1/\sqrt{6}, -1/\sqrt{6}, -1), (- \sqrt{2/3}, 1/\sqrt{6}, 1/\sqrt{6}, -1)$ and their permutations. Other interesting duality-related combinations are $(-1/3, 2/3, 2/3, 0)$, $(1/3, -2/3, -2/3, -2)$ and $(0, 1/\sqrt{2}, - 1/\sqrt{2}, -1), (0, -1/\sqrt{2}, 1/\sqrt{2}, -1)$. The last two give the so-called Taub points (flat Minkowski spacetime) \cite{wainwright}. 

The question arises whether the changes of the Kasner indices (chaotic and duality-related) have something in common. In order to answer this question one should try to find a suitable parametrization of the Kasner indices analogous  to the one given in \cite{belkhal} for the stiff-fluid case. We have checked that such a parametrization does not cover the right range of the indices given by (3.58) unless we multiply $p_1$ by $\sqrt{3}$. After some searching, we have found the following parametrization of the indices which has advantages which we will explain in due course. 

The $u$-parametrization we use is defined as follows \footnote{In the general relativity case $(M=0)$ this parametrization can be written down in terms of the one suggested in \cite{wainwright} with $p_1 = (1/3)(1 - 2 \cos{\psi}), p_{2,3} = (1/3)(1 + \cos{\psi} \pm \sin{\psi})$ with a suitable choice of $\cos{\psi} = (1/2)(u^2 - 6u + 1)(u^2 + 1)$ in our case.}

\bea
\bar{p}_1 & = & \frac{2u}{1 + u^2}   ,\\
\bar{p}_2 & = & \frac{1}{2} \frac{1}{1 + u^2} \left[ (1 + M)(1 + u^2) - 2u + P^{+}(M,u) \right]   \nonumber ,\\
\bar{p}_3 & = & \frac{1}{2} \frac{1}{1 + u^2} \left[ (1 + M)(1 + u^2) - 2u - P^{+}(M,u)   \right] \nonumber,
\eea
where $u$ is constant and 
\be
P^{+}(M,u) = \sqrt{\left(1 - 2M - M^2\right) \left(1 + u^2 \right)^2 
 + 4u \left(M + 1\right) \left(1 + u^2\right) - 12u^2}  ,
\ee
and to achieve $p_1 < 0$ we need $u < 0$. Alternatively, we have
\bea
p_1 & = & - \frac{2u}{1 + u^2}   ,\\
p_2 & = & \frac{1}{2} \frac{1}{1 + u^2} \left[ (1 + M)(1 + u^2) + 2u + P^{-}(M,u) \right]   \nonumber ,\\
p_3 & = & \frac{1}{2} \frac{1}{1 + u^2} \left[ (1 + M)(1 + u^2) + 2u - P^{-}(M,u)   \right] \nonumber,
\eea
where
\be
P^{-}(M,u) = \sqrt{\left(1 - 2M - M^2\right) \left(1 + u^2 \right)^2 
 - 4u \left(M + 1\right) \left(1 + u^2\right) - 12u^2}  ,
\ee
and for $p_1 < 0$ we need $u > 0$. These are very general transformations that  include all duality-related cases (3.58)-(3.59) and the vacuum general relativity  case when $M = 0$. Similarly, as for stiff-fluid models in Ref.\cite{belkhal}, we draw a plot the allowed values of the parameter $u$ against $M$ and show the duality-related regions in Figs.1 and 2.

There are many possible transformations of the general homographic type $u \to (au + b)/(cu + d)$ or some more general types related to Pad\'e or other rational approximants \cite{pade} which may be useful in the discussion of Mixmaster oscillations. First, notice that a transformation of the type 
\be
u \to \frac{1}{u}   ,
\ee
gives
\bea
\bar{p}_i(M,1/u) & = & \bar{p}_i(M,u)   ,\\
p_i(M,1/u) & = & p_i(M,u)   \nonumber,
\eea
where $i = 1,2,3$, unless we take negative value of the root of $P^{\pm}$ after applying the transformation. On the other hand using 
\be
u \to - \frac{1}{u}   ,
\ee
changes parametrization (3.60) into (3.62), i.e.,
\bea
\bar{p_i}(M,-1/u) & = & p_i(M,u) = - \bar{p}_i(M,u)   ,\\
p_i(M,-1/u) & = & \bar{p}_i(M,u) = - p_i(M,u)   .
\eea

Second, one can consider the transformation 
\be
u \to - u   ,
\ee
which results in the same rules as (3.65). However, this transformation is not enough to describe duality relations between the considered solutions. As one can see, the duality-reflecting transformations should be of the type (see also Fig. 1) 
\be
u \to -u, \hspace{0.3cm} M \to - (M + 2)   ,
\ee
or, alternatively
\be
u \to - \frac{1}{u}, \hspace{0.3cm} M \to - (M + 2)   .
\ee 
From (3.70) and (3.71) we see that 
\be
P^{\pm}(- M - 2, - u) = P^{\pm}(- M - 2, - 1/u) = P^{\pm}(M, u)   ,
\ee
and 
\bea
\bar{p}_1(-u) & = & - \bar{p}_1(u)   ,\\
\bar{p}_2(- M - 2, - u) & = & - \bar{p}_3(M, u)   \nonumber, \\
\bar{p}_3(- M - 2, - u) & = & - \bar{p}_2(M, u)   \nonumber; 
\eea
similarly,
\bea
p_1(-u) & = & - p_1(u)   ,\\
p_2(- M - 2, - u) & = & - p_3(M, u)  \nonumber, \\
p_3(- M - 2, - u) & = & - p_2(M, u)  \nonumber.
\eea
We notice that the duality in (3.73)-(3.74) would entail the exchange of the Kasner indices $p_2$ and $p_3$.

If we assume that $a \gg b \gg c$ in (3.23)-(3.24) then we obtain the
following set of approximating equations (for $A = 0$)
\bea
\alpha_{, \eta \eta} & = & - \frac{1}{2} e^{4\alpha} e^{-2\Lambda M\eta}  \nonumber,\\
\beta_{, \eta \eta} & = &  \frac{1}{2} e^{4\alpha} e^{-2\Lambda M\eta}   ,\\
\gamma_{, \eta \eta} & = &  \frac{1}{2} e^{4\alpha} e^{-2\Lambda M\eta}  \nonumber,\\
\phi_{, \eta \eta} & = & 0   \nonumber,
\eea
together with the constraint (3.19). The Kasner solutions in terms of 
$\eta$-time are given by
\bea
\alpha(\eta) & = & \Lambda p_1 \eta + {\rm const}   \nonumber,\\
\beta(\eta) & = & \Lambda p_2 \eta + {\rm const}   \nonumber,\\
\gamma(\eta) & = & \Lambda p_3 \eta + {\rm const}   ,\\
\phi(\eta) & = &  \Lambda M \eta + {\rm const.}   \nonumber
\eea

However, these solutions are not directly obtained by using the relations
(3.18) between the Einstein frame and string-frame scale factors. This is
very important in the case when the axion field is taken into account and will be discussed below.

The solutions of equations (3.75) which fulfil the above conditions, 
(3.76), in the limit $\eta \to \infty$ ($t \rightarrow \infty$ - i.e. far from singularity), provided $p_1  = - \vert p_1 \vert < 0$ and $2p_1 - M = -2 \vert p_1 \vert - M < 0$, 
 can be chosen to be\footnote{Note we can derive them in a similar way to the 
for $A \neq 0$ case (see the rest of this section), but here it is more 
convenient to follow the results given in \cite{ishihara}.}
\bea
\alpha(\eta) & = & -\frac{1}{2} \ln{\left(\frac{1}{\Lambda (2p_1 - M)} 
\cosh{\Lambda (2p_1 - M)\eta}\right)} + \frac{1}{2} \Lambda M \eta   \nonumber,\\
\beta(\eta) & = & \frac{1}{2} \ln{\left(\frac{1}{\Lambda (2p_1 - M)} 
\cosh{\Lambda (2p_1 - M)\eta}\right)} + \left(p_1 + p_2 - M \right) \Lambda 
\eta + \frac{1}{2} \Lambda M \eta   ,\\
\gamma(\eta) & = & \frac{1}{2} \ln{\left(\frac{1}{\Lambda (2p_1 - M)} 
\cosh{\Lambda (2p_1 - M)\eta}\right)} + \left(p_1 + p_3 - M \right)\Lambda \eta + 
\frac{1}{2} \Lambda M \eta   \nonumber.
\eea

In the limit $\eta \to - \infty$ ($t \to 0$ - i.e. on the approach of singularity) they approach the following asymptotic forms 
\footnote{If we assume $p_1 < 0$ and $2p_1 - M < 0$, then $p_1 - M < 
0$, provided $M > 0$, which means we have changed expansion of $a(\eta)$ into 
contraction.}
\bea
\alpha(\eta) & \sim & - \Lambda (p_1 - M) \eta   \nonumber,\\
\beta(\eta) & \sim & \Lambda (p_2 + 2p_1 - M) \eta   ,\\
\gamma(\eta) & \sim & \Lambda (p_3 + 2p_1 - M) \eta   \nonumber,\\
\phi(\eta) & \sim & \Lambda M \eta \nonumber.
\eea
One can check the solutions (3.77) by putting them into the constraint (3.19) 
in order to recover the condition (3.54), as expected.

Now, we can express the scale factors in terms of the new Kasner parameters
\bea
a & = & a_0^{\prime} t^{p_1^{\prime}}   \nonumber,\\
b & = & b_0^{\prime} t^{p_2^{\prime}}   ,\\
c & = & c_0^{\prime} t^{p_3^{\prime}}   \nonumber,\\
e^{-\phi} & = & d_0^{\prime} t^{p_4^{\prime}}   \nonumber,
\eea
where
\bea
p_1^{\prime} & = & - \frac{p_1 - M}{1 + 2p_1 - M}   \nonumber,\\
p_2^{\prime} & = & \frac{p_2 + 2p_1 - M}{1 + 2p_1 - M}   \nonumber,\\
p_3^{\prime} & = & \frac{p_3 + 2p_1 - M}{1 + 2p_1 - M}   ,\\
p_4^{\prime} & = & \frac{- M}{1 + 2p_1 - M} = - M^{\prime}  \nonumber,
\eea
and 
\bea
\Lambda^{\prime} & = & a_0^{\prime}b_0^{\prime}c_0^{\prime}d_0^{\prime}
\nonumber ,\\
\eta & = & (\Lambda^{\prime})^{-1} \ln{t} + {\rm const.}   ,\\
\Lambda^{\prime} & = & (1 + 2p_1 - M) \Lambda   \nonumber.
\eea

If we take the axion field into account ($A \neq 0$) and assume that $a \gg b \gg c$ in
(3.23)-(3.25)), then we obtain\footnote{If $M > 0$, then the term $e^{-2\phi}$
increases for $\eta \to - \infty$. If, in turn, 
$p_1 < 0$, $M > 0$, and $2p_1 - M > 0$, then the term $e^{2(2p_1 - M)\eta}$
decreases for $\eta \to -\infty$ and the whole picture is dominated by the axion
term $1/2 A^2 e^{-2\phi}$. It follows that the field equations become isotropic 
$\alpha_{, \eta \eta} = \beta_{, \eta \eta} = \gamma_{\prime \eta
\eta} = 1/2 \phi_{, \eta \eta} = 1/2 A^2 e^{-2\phi}$. The conclusion is
that axion isotropises the model and chaos is impossible in
such a case. It seems that the time-dependent axion field ansatz (cf. Appendix 
A) would allow chaos, but it is not admitted by the BIX geometry.}
\bea
\alpha_{, \eta \eta} & = & \frac{1}{2} \left( A^2 - e^{4\alpha} \right) 
e^{-2\phi}   \nonumber,\\
\beta_{, \eta \eta} & = & \frac{1}{2} \left( A^2 + e^{4\alpha} \right) 
e^{-2\phi}   \nonumber,\\
\gamma_{, \eta \eta} & = & \frac{1}{2} \left( A^2 + e^{4\alpha} \right) 
e^{-2\phi}   ,\\
\phi_{, \eta \eta} & = & A^2 e^{-2\phi}   \nonumber,
\eea
with $\phi(\eta)$ given by (3.14). The Kasner solutions in terms of 
$\eta$-time are now given by (compare (3.18))
\bea
\alpha(\eta) & = & \frac{\Lambda}{2} \left(p_1 + M \right) \eta + {\rm const}
\equiv \Lambda q_1 \eta + {\rm const.}   \nonumber,\\
\beta(\eta) & = & \frac{\Lambda}{2} \left(p_2  + M \right) \eta + {\rm const}   
\equiv \Lambda q_1 \eta + {\rm const.}     \nonumber,\\
\gamma(\eta) & = & \frac{\Lambda}{2} \left(p_3 + M \right) \eta + {\rm const}   
\equiv \Lambda q_1 \eta + {\rm const.}     ,\\
\phi(\eta) & = & \Lambda M \eta + {\rm const.}   \nonumber
\eea

The solutions which fulfil the above initial conditions (3.83) for 
$\eta \to \infty$ ($p_1 < 0$) are 
\bea
\alpha(\eta) & = & - \frac{1}{2} \ln{\left( \frac{1}{p_1} \cosh{p_1 \eta} \right)} 
+ \frac{1}{2} \phi(\eta)   ,\\
\beta(\eta) & = &  \frac{1}{2} \ln{\left( \frac{1}{p_1} \cosh{p_1 \eta} \right)} 
+ \frac{1}{2} \left( p_1 + p_2 \right) \eta + \frac{1}{2} \phi(\eta)   ,\\
\gamma(\eta) & = &  \frac{1}{2} \ln{\left( \frac{1}{p_1} \cosh{p_1 \eta} \right)} 
+ \frac{1}{2} \left( p_1 + p_3 \right) \eta + \frac{1}{2} \phi(\eta)   ,\\
\phi(\eta) & = & \ln{\left[ \cosh {M \eta }+
\sqrt{1-\frac{A^2}{M^2}}\sinh {M\eta}\right] }   ,
\eea
or, alternatively
\bea
\alpha(\eta) & = & - \frac{1}{2} \ln{\left( \frac{1}{2q_1 - M} \cosh{(2q_1 - M)
\eta} \right)} + \frac{1}{2} \phi(\eta)   ,\\
\beta(\eta) & = &  \frac{1}{2} \ln{\left( \frac{1}{2q_1 - M} 
\cosh{(2q_1 - M) \eta} \right)} 
+ \left( q_1 + q_2 - M \right) \eta + \frac{1}{2} \phi(\eta)   ,\\
\gamma(\eta) & = &  \frac{1}{2} \ln{\left( \frac{1}{2q_1 - M} 
\cosh{(2q_1 - M)\eta} \right)} 
+ \left( q_1 + q_3 - M \right) \eta + \frac{1}{2} \phi(\eta)   ,
\eea
with $\phi(\eta)$ unchanged. 

One can easily check by putting these solutions into the constraint (3.19), 
that the condition (3.54) is fulfilled, which in turn ensures that the conditions
(3.50)-(3.51) are fulfilled. In particular, note that, for (3.88)-(3.90), we need to replace $p_i's$ by $q_i's$. In the limit $\eta \to - \infty$, (that is $t \to 0$), they approach 
the following forms 
\bea
\alpha(\eta) & \sim & - \frac{\Lambda}{2} (p_1 - M) \eta   \nonumber,\\
\beta(\eta) & \sim & \frac{\Lambda}{2} (p_2 + 2p_1 - M) \eta   \nonumber,\\
\gamma(\eta) & \sim & \frac{\Lambda}{2} (p_3 + 2p_1 - M) \eta   ,\\
\phi(\eta) & \sim & - \frac{\Lambda}{2} M \eta \nonumber,
\eea
or
\bea
\alpha(\eta) & \sim & - \Lambda (q_1 - M) \eta   \nonumber,\\
\beta(\eta) & \sim & \Lambda (q_2 + 2q_1 - M) \eta   \nonumber,\\
\gamma(\eta) & \sim & \Lambda (q_3 + 2p_1 - M) \eta   ,\\
\phi(\eta) & \sim & - \Lambda M \eta \nonumber.
\eea
Having given the conditions (3.50)-(3.51), one can express the indices 
$p_2$ and $p_3$ by using $p_1$ and $M$, i.e.,
\bea
p_2 = \frac{1}{2} \left[ \left( M + 1 - p_1 \right) - \sqrt{ - 3p_1^2 + 
2p_1 \left(M + 1\right) + 1 - M\left(M + 2\right)} \right]  \nonumber,\\
p_3 = \frac{1}{2} \left[ \left( M + 1 - p_1 \right) + \sqrt{ - 3p_1^2 + 
2p_1 \left(M + 1\right) + 1 - M\left(M + 2\right)} \right]  .
\eea
Since the expression under the square root should be nonnegative, one can extract the restriction (3.56) on the permissible
values of $M$. However, we are interested in knowing whether the curvature
terms on the right-hand side of the field equations (3.23)-(3.25) really increase
as $\eta \to - \infty$ ($t \to 0$). This would require either $a^4e^{-2\phi}, b^4e^{-2\phi}$, or 
$c^4e^{-2\phi}$ to increase if the transition to another Kasner epoch is to 
occur \cite{halp,ishihara}. Since 
\bea
a^4e^{-2\phi} \propto t^{(2p_1 - M)} = t^{(1 + p_1 - p_2 - p_3)}   \nonumber,\\
b^4e^{-2\phi} \propto t^{(2p_2 - M)} = t^{(1 + p_2 - p_3 - p_1)}   ,\\
c^4e^{-2\phi} \propto t^{(2p_3 - M)} = t^{(1 + p_3 - p_1 - p_2)}   \nonumber,
\eea
we need one of the following three conditions to be fulfilled (remember that we have assumed $p_1 < 0, M > 0$):
\bea
2 p_1 - M & = & 1 + p_1 - p_2 - p_3  <  0   \nonumber,\\
2 p_2 - M & = & 1 + p_2 - p_3 - p_1  <  0   ,\\
2 p_3 - M & = & 1 + p_3 - p_1 - p_2  <  0   \nonumber.
\eea
The three conditions (3.95), with the help of (3.93), are equivalent to 
\bea
p_1 & < & \frac{M}{2}   ,\\
-3p_1^2 + 2p_1\left( M + 1 \right) + 1 - M \left( M + 2 \right) & > & 0.
\eea
A plot of these conditions is given in Fig. 3.
The last of these conditions, (3.97), provides bounds on the
possible values of $M$ if a transition to occur:
\be
-2 \leq M \leq \frac{2}{3}   .
\ee
Now, we see that the regions where the Friedmann isotropic limit is possible (all the
Kasner indices equal - this happens for $M = -1 - \sqrt{3}$ and $M = -1 +
\sqrt{3}$) are excluded. One can always find the range of the
indices for a transition from one Kasner epoch to another to occur in the
string frame.

Instead of expressing the conditions for Kasner-type transitions in terms of
$p_1$ and $M$, we can follow the pattern of \cite{halp} and write them in
terms of $p_1$ and $p_2$. From the conditions (3.50)-(3.51), we can write 
\bea
p_3 & = & M + 1 - p_1 - p_2    ,\\
M & = & p_1 + p_2 - 1 \pm \sqrt{1 - p_1^2 - p_2^2}   ,
\eea
where the plus sign is for $M > - 1, p_3 > 0$ and minus sign for $M < - 1, p_3 < 0$.
So, $p_1$ and $p_2$ must be such that ($M$ real) 
\be
1 - p_1^2 - p_2^2 \ge 0   ,
\ee
and one of the three conditions (3.95) must be fulfilled, i.e., either 
\bea
p_1^2 + p_2^2 - p_1p_2 + p_1 - p_2 & < & 0   \nonumber,\\
p_1^2 + p_2^2 - p_1p_2 - p_1 + p_2 & < & 0   ,\\
p_1^2 + p_2^2 + p_1p_2 - p_1 - p_2 & > & 0   \nonumber,
\eea
for $p_3 > 0$, or
\bea
p_1^2 + p_2^2 - p_1p_2 + p_1 - p_2 & < & 0   \nonumber,\\
p_1^2 + p_2^2 - p_1p_2 - p_1 + p_2 & < & 0   ,\\
p_1^2 + p_2^2 + p_1p_2 + p_1 + p_2 & > & 0   \nonumber,
\eea
for $p_3 <  0$.

The plot of these conditions is given in Fig.4.

In summary, we have been able to determine the range of values that can be taken by the Kasner indices in the string frame and have proposed a parametrization which describes the evolution of these indices. Also, we have determined the values of the Kasner indices (see Figs. 3 and 4) for which the spacetime  oscillations can really take place. However, now we have to determine whether the oscillations can be stopped once they have started as $t \to 0$. Before we come to this in Section V we first discuss some relations between the Kasner indices and duality.

\section{Exchange of Kasner Indices and Duality}
\setcounter{equation}{0}

The class of homogeneous solutions of the low-energy-effective-action equations of string cosmology have a
global $O(3,3)$ invariance also called $T$-duality (other duality invariance  called $S$-duality also appears in superstring models, see e.g. \cite{bframe}) under which ($\bar{\phi}$ is the so-called shifted
dilaton field)
\be
{\bf M} \to {\bf M}^{\prime} = {\bf \Omega}^{T} {\bf M} {\bf \Omega}   ,\hspace{2.cm} \bar{\phi} \equiv \phi
- \ln{\sqrt{{\rm det {\bf G}}}} \to \bar{\phi}   ,
\ee
where ${\bf \Omega}$ is $6 \times 6$ constant matrix satisfying 
\be
{\bf \Omega}^{T} {\bf \Pi} {\bf \Omega} = {\bf \Pi} , \hspace{2.cm}
\ {\bf \Pi} \ = 
{{\bf 0} \;\;  {\bf 1} 
\choose
{\bf 1} \;\;  {\bf 0}}   ,
\ee
and {\bf 1} is the $3 \times 3$ identity matrix and 
\be 
{\bf M} \equiv 
{ {\bf G}^{-1} \;\; -{\bf G}^{-1}{\bf B} 
\choose
{\bf BG}^{-1}  \;\; {\bf G} - {\bf BG}^{-1}{\bf B}},
\ee
where ${\bf G} = g_{ij}$ and ${\bf B} = B_{ij}$ are $3 \times 3$ matrices. Any $6 \times 6$
constant matrix ${\bf \Omega}$ obeying Eq.(5.2) generates new solutions ${\bf M}^{\prime}$
from the original set ${\bf M}$. If the antisymmetric tensor field (axion) vanishes
under a choice ${\bf \Omega} = {\bf \Pi}$, we deal with the "scale factor duality" where
\bea
a^{\prime 2} & \to & \frac{1}{a^2}   \nonumber,\\
b^{\prime 2} & \to & \frac{1}{b^2}   ,\\
c^{\prime 2} & \to & \frac{1}{c^2}   \nonumber,\\
\phi^{\prime} & \to & \phi - 2 \ln{abc}   \nonumber. 
\eea

It is useful to define the logarithm of an average scale factor $\bar{\beta}$ and the so-called shifted dilaton $\bar{\phi}$ defined by \cite{gv}
\bea
\bar{\beta_i} & = & \frac{1}{\sqrt{3}} \ln{a_i}    ,\\
\bar{\beta} & = & \frac{1}{\sqrt{3}} \ln{(abc)}  \nonumber ,\\
\bar{\phi} & = & \phi - \sqrt{3} \bar{\beta}   \nonumber.
\eea
Using (3.40)-(3.45) we have the relations (4.5) in terms of Kasner indices, i.e.,
\bea
\bar{\phi} & = & \phi - \ln{\Lambda} - (M + 1) \ln{t}   \nonumber,\\
\bar{\beta}_i & = & \frac{1}{\sqrt{3}} \left(\ln{a_{0i}} + p_i \ln{t} \right)  ,\\
\bar{\beta} & = & \frac{1}{\sqrt{3}} \left[\ln{\Lambda} + (M + 1) \ln{t} \right]   \nonumber,
\eea
where $a_{0i} = \{a_0, b_0, c_0\}$. In these variables the duality symmetry is just expressed by 
\bea
\bar{\beta_i}(t) & = & - \bar{\beta_i}(t)   ,\\
\bar{\beta}(t) & = & - \bar{\beta}(t)   \nonumber,\\
\bar{\phi}(t) & = & \bar{\phi}(t)   \nonumber;
\eea
or, in terms of Kasner indices, it reads $p_i \to - p_i, M + 1 \to - (M + 1)$. 
After inclusion of time symmetry we have \cite{gv}
\bea
\bar{\beta_i}(t) & = & - \bar{\beta_i}(-t)   ,\\
\bar{\beta}(t) & = & - \bar{\beta}(-t)   \nonumber,\\
\bar{\phi}(t) & = & \bar{\phi}(-t)   \nonumber.
\eea
In the isotropic case $p_i = \pm 1/\sqrt{3}$ and we recover exactly the
case given in Ref.\cite{gv}.

The same relations can be written down using the time coordinate $\eta$ instead of $t$. 
Using the exact expressions, (3.76), for Kasner solutions we have 
\bea
\bar{\beta} & = & \frac{\Lambda}{\sqrt{3}} (M + 1) \eta   \nonumber,\\
\bar{\phi} & = & \phi - \sqrt{3} \bar{\beta} - \Lambda \eta   \nonumber,\\
\bar{\beta}_i & = & \frac{\Lambda}{\sqrt{3}} p_i \eta   .
\eea
So, one could relate these by duality symmetries
\bea
\bar{\phi}(\eta) \to \bar{\phi}(-\eta)  ,\\
\bar{\beta}(\eta) \to - \bar{\beta}(-\eta)  ,
\eea
for $\eta \to \pm \infty$ respectively, using chaotic changes $p_i \to - p_i$.

\section{Hamiltonian Approach to BIX String Models}

In this Section we formulate a generalized Kasner model in Hamiltonian formalism as in Refs. \cite{ryan,halp,hosoya} in order to discuss the conditions for an infinite sequence of scatterings to occur against the walls of the curvature potential. As in the previous sections we discuss the problem in both the Einstein and the string frames. We also introduce the so-called B-frame, or axion frame \cite{bframe}, in which axion is minimally coupled. 

\subsection{Einstein Frame}

We introduce the following standard parametrization for the Einstein-frame scale factors 
\bea
\tilde{a} & = & e^{\tilde{\alpha} + \psi_+ + \sqrt{3} \psi_-}   \nonumber ,\\
\tilde{b} & = & e^{\tilde{\alpha} + \psi_+ - \sqrt{3} \psi_-}   \nonumber ,\\
\tilde{c} & = & e^{\tilde{\alpha} - 2\psi_-}   ,
\eea
and we define the potential, which describes the spatial curvature anisotropy (2.21) felt by scale factors in the Einstein-frame by  
\be
\tilde{V}(\psi_{\pm}) = e^{-2\tilde{\alpha}} V(\psi_{\pm})   ,
\ee
where
\be
V(\psi_{\pm}) = \frac{1}{2} \left[ e^{-8 \psi_{+}} + 2 e^{4\psi_{+}} \left( \cosh{4 \sqrt{3} \psi_{-}} - 1 \right) - 4 e^{-2\psi_{+}} \cosh{2 \sqrt{3} \psi_{-}} \right]   .
\ee
Using (5.1)-(5.2), the Eqs. (3.7)-(3.10) read as 
\be
\tilde{\alpha}^{\prime 2} = \psi_{+}^{\prime 2} + \psi_{-}^{\prime 2} + \frac{1}{12} \phi^{\prime 2} + 
\frac{1}{12} A^2 e^{-2\phi - 6\tilde{\alpha}} + \frac{1}{6} e^{-2\tilde{\alpha}} V(\psi_{\pm})   .
\ee
This is the Hamiltonian constraint. The Einstein-frame action in terms of the scale factors (5.1) after integrating out spatial variables is given by (compare \cite{james}) 
\be
S = \int{d\tilde{t} e^{3\tilde{\alpha}} \left[ -6 \tilde{\alpha}^{\prime 2} + 6\psi_{+}^{\prime 2} + 6\psi_{-}^{\prime 2} + \frac{1}{2} \phi^{\prime 2} + 
\frac{1}{2} A^2 e^{2\phi} \sigma^{\prime 2}  +  e^{-2\tilde{\alpha}} V(\psi_{\pm}) \right]}    ,
\ee
and the conjugate momenta are 
\bea
\pi_{\tilde{\alpha}} & = & - 12 \tilde{\alpha}^{\prime} e^{3\tilde{\alpha}}   \nonumber,\\
\pi_{+} & = & 12 \psi_{+}^{\prime} e^{3\tilde{\alpha}}   \nonumber,\\
\pi_{-} & = & 12 \psi_{-}^{\prime} e^{3\tilde{\alpha}}   \nonumber,\\
\pi_{\phi} & = & \phi^{\prime} e^{3\tilde{\alpha}}   \nonumber,\\
\pi_{\sigma} & = &  \sigma^{\prime} e^{2 \phi + 3\tilde{\alpha}} = \rm{const.} = A    ,
\eea 
so the Hamiltonian is 
\be
H = - \frac{\pi_{\alpha}^2}{24} + \frac{\pi_+^2}{24} + \frac{\pi_-^2}{24} + \frac{\pi_{\phi}^2}{2} +  \frac{\pi_{\sigma}^2}{2} + e^{4\tilde{\alpha}} V(\psi_{\pm})   .
\ee
Now, we follow the standard discussion of the potential walls 
\be
\bar{V} = e^{4\tilde{\alpha}} V(\psi_{\pm})    
\ee
being hit by a particle moving in the potential well (see Ref.\cite{hosoya}). In the region $\psi_{+} \ll - 1$ and $\psi_-  \approx 0$, the approximate distance from the origin of coordinates $\psi_+$ and $\psi_-$ to the wall is given by 
\be
D = - \frac{1}{2} \tilde{\alpha}   ,
\ee
while the maximum apparent velocity of this wall is 
\be
v_{max} = \tilde{\alpha}^{\prime}   .
\ee
The velocity of a particle moving against the walls is 
\be
v_p = \sqrt{\psi_+^{\prime 2} + \psi_-^{\prime 2}}    ,
\ee
and it will not be scattered infinitely many times if there is some region of the potential which the particle enters and from which it cannot catch up with the wall, i.e., if
\be
v_p =  \sqrt{\psi_+^{\prime 2} + \psi_-^{\prime 2}}  < \tilde{\alpha}^{\prime} \approx \sqrt{ \psi_+^{\prime 2} + \psi_-^{\prime 2} + (1/12) \phi^{\prime 2} + (1/12) A^2 e^{-2\phi - 6 \tilde{\alpha}} }   .
\ee
Clearly, this condition is fulfilled in every case unless $\phi^{\prime} = A = 0$ (no dilaton and axion - that is, the general relativity vacuum regime), which reflects the fact that a particle cannot be scattered infinitely many times and that {\em there is no chaos in the Einstein frame}. This result is expected since in the Einstein frame both dilaton and axion fields behave as stiff fluids with the equation of state $p = \varrho$. A numerical dicussion of 10-dimensional axion-dilaton low-energy effective-action models in the Einstein frame where nine dimensions were split into three isotropic 3-dimensional spaces leading effectively to our anisotropic 4-dimensional model, was also given in Ref.\cite{andrew} with the same final conclusion about the non-existence of chaos. 

\subsection{String Frame and Axion frame}

In the string frame we can use the same parametrization as in (5.1), but we just drop the tildes. The potential (5.2) can also be used without tildes. In that parametrization Eq.(3.5) becomes  
\be 
\dot{\alpha}^2 =\dot{\psi}_+^2  + \dot{\psi}_-^2 - \frac{1}{6} \dot{\phi}^2 + \frac{1}{6} e^{-2 \alpha} V(\psi_{\pm}) + \frac{1}{12} A^2 e^{-6 \alpha} + \dot{\alpha}\dot{\phi}   .
\ee
After applying the variables $\bar{\beta}$ and $\bar{\phi}$ defined by (4.5) we  can remove the $\dot{\phi} \dot{\alpha}$ term, obtaining 
\be
\dot{\bar{\phi}}^2 = \dot{\bar{\beta}}^2 + 6 \dot{\psi}_+^2  + 6 \dot{\psi}_-^2 + e^{-2 \frac{\bar{\beta}}{\sqrt{3}}} V(\psi_{\pm}) + \frac{1}{2} A^2 e^{- 2 \sqrt{3} \bar{\beta}}   .
\ee
Following the analysis given in \cite{gv}, we apply a new time coordinate $\tau$ defined as 
\be
dt = d\tau e^{-\bar{\phi}}   ,
\ee
and define a new variable, $y$, which is the logarithm of an averaged scale factor in the conformally related axion frame (or B-frame) in which the axion is minimally coupled \cite{bframe}. This is given by 
\be
y \equiv \sqrt{3} \bar{\phi} + \bar{\beta} = \frac{1}{\sqrt{3}} \ln{abc}   ,
\ee
and brings (5.14) to the form ($(...)_{\tau} = d(...)/d\tau$) 
\be
y_{\tau}^2 = \phi_{\tau}^2 + 12 \psi_{+\tau}^2 + 12 \psi_{-\tau}^2 + A^2 e^{- 2\phi} + 2 e^{- \frac{2}{\sqrt{3}} y} V(\psi_{\pm})   .
\ee
Eq. (5.17) is, in fact, the Hamiltonian constraint obtained from the action $(\lambda_s^2 = 8\pi G)$
\be
S = \frac{\lambda_s}{4} \int{d\tau \left[ \phi_{\tau}^2 - y_{\tau}^2 + 12 \left( \psi_{+\tau}^2 + \psi_{-\tau}^2 \right) - A^2 e^{-2 \phi} - 2 e^{- \frac{2}{\sqrt{3}} y} V(\psi_{\pm}) \right] }    .
\ee
The canonical momenta are then
\bea
\pi_{\phi} & = & \frac{\lambda_{s}}{2} \phi_{\tau}   \nonumber,\\
\pi_{y} & = & - \frac{\lambda_{s}}{2} y_{\tau}   \nonumber,\\
\pi_{+} & = & 6 \lambda_{s} \psi_{+\tau}   \nonumber,\\
\pi_{-} & = & 6 \lambda_{s} \psi_{-\tau}   ,
\eea
and the Hamiltonian is just  
\be
H = \frac{1}{\lambda_{s}} \left[ \pi_{\phi}^{2} - \pi_{y}^{2} + \frac{1}{12} \left( \pi_{+}^{2} + \pi_{-}^{2} \right) + \frac{\lambda_{s}^{2}}{4} A^2 e^{-2\phi} + \frac{\lambda_{s}^{2}}{2} e^{-\frac{2}{\sqrt{3}}y} V(\psi^{\pm}) \right]   .
\ee
Following the analysis of the previous subsection V.A., we see that the maximum apparent velocity of the wall at $\psi_{+} \ll - 1, \psi_- \approx 0$ is given by 
\be
v_{max} = \frac{1}{2\sqrt{3}} y_{\tau}   ,
\ee
and the condition for chaotic scatterings to cease is just that 
\be
v_p = \sqrt{\psi_{+\tau}^2 + \psi_{-\tau}^2} < \frac{1}{2\sqrt{3}} y_{\tau} \approx 
 \sqrt{\psi_{+\tau}^2 + \psi_{-\tau}^2 + \frac{1}{12} \phi_{\tau}^2 + \frac{1}{12} A^2 e^{-2 \phi} }   ,
\ee
This is clearly fulfilled except in the general relativity case where $\phi = A = 0$ (i.e., no axion and dilaton fields). This gives our final conclusion that {\em there is no chaos in BIX string cosmology in the string or axion frames}.

It is not surprising that the physical behaviour should be similar in every frame \cite{veldom}, so that if there is no chaos in the Einstein frame there should not be chaos in any other frame. String theory appears to impose too much symmetry through its duality invariances for chaos to appear \footnote{Chaos has been studied in other related situations in Ref.\cite{Mtheory}}.

\section{Discussion}

In this paper we have carried out a detailed analysis of the spatially
homogeneous universes of Bianchi type IX in the context of low-energy string
theory. These universes are of special interest in relativistic cosmology
because they display chaotic behaviour in vacuum and in the presence of
fluids with $p<\rho $. They were originally termed 'Mixmaster' universes by
Misner because they offered the possibility for light to travel all the way
around the universe in different directions. Moreover, they are the most
general closed universes which are spatially homogeneous and may be closely
related to parts of the general solution of Einstein's equations in the
neighbourhood of a strong curvature singularity of the sort that
characterises the initial state of general relativistic cosmological models.
This behaviour has also been extensively investigated because it is of
intrinsic mathematical interest. It is also known that its occurrence in
general relativity depends upon the dimensionality of space. We investigated
the string cosmological equations for the type IX metric. We found that
chaotic behaviour does not occur in string cosmology in either the Einstein
or the string or axion frames. While it is possible for finite sequences of
oscillations to occur in the scale factors' evolution on approach to $t=0$,
these oscillations cannot continue indefinitely. They inevitably terminate
in a state in which all the three orthogonal scale factors decrease with
decreasing time monotonically on approach to the initial singularity. We
investigated the detailed sequences of evolutionary changes that can take
place in the evolution during the finite sequences of oscillations between
epochs which are well approximated by Kasner universes. We found that the
duality symmetry required of the string evolution introduced new invariances
for the possible changes in the Kasner parameters in addition to those which
characterise the Kasner-to-Kasner cycles of oscillations.  The requirements
of duality invariance on the evolution of the metric appear to be so
constraining that chaotic behaviour is excluded. We have obtained these
results in two complementary ways: by direct matching of asymptotic
expansions to the solutions of the sytem of non-linear ordinary differential
equations of string cosmology and by use of the Hamiltonian formulation of
cosmology. In the Hamiltonian picture the evolution of the type IX string
cosmology is represented as the motion of a 'universe point' inside a
potential that is open only along three narrow channels. The walls of this
'almost closed' potential of H\'enon-Heiles type expand outwards as the
singularity is reached. Whereas in the vacuum models of general relativity,
the universe point always catches the walls and bounces chaotically around
within the potential, in string theory the universe point need never catch
the walls. If it is moving towards a wall at a very oblique angle then the
normal component of its velocity towards the wall can become too small for
it ever to catch the wall. In general, we find that this situation always
arises after a finite number of collisions have occurred in the Mixmaster
string cosmology. The resulting asymptotic state is therefore similar to
that in a model with no potential walls at all; that is the Bianchi type I
or Kasner universe.

\section{Acknowledgments}

MPD wishes to thank Amithaba Lahiri and David Wands for useful discussions,
Malcolm MacCallum for helpful e-mail correspondence and the NATO/Royal Society for the 
support while at the University of Sussex. MPD was also partially supported by the 
Polish Research Committee (KBN) grant No 2 PO3B 196 10. JDB is supported 
by PPARC.

\appendix

\section{Ricci tensor in a coordinate frame}

\setcounter{equation}{0}

In this Appendix we discuss a possibility of admittinig a time-dependent antisymmetric tensor potential $B_{\mu\nu} = B_{\mu\nu}(t)$ as given in Ref. \cite{ed} to the axisymmetric Bianchi IX model. It has been proven \cite{homog} that such a potential cannot be admitted to a general (i.e. non-axisymmetric case).  However, to achieve this, we first give the components of Ricci tensor for the axisymmetric Bianchi IX model in terms of coordinates rather than in orthonormal frames of Section II. For the sake of generality, we start with a general metric.

The metric (2.9) in a coordinate frame has the following components 
\bea
g_{\0\0} & = & 1   ,\\
g_{\1\1} & = & - c^2(t)   ,\\
g_{\2\2} & = & - \left[ a^2(t) \cos^2{\psi} + b^2(t) \sin^2{\psi} \right]   ,\\
g_{\3\3} & = & - \sin^2{\theta} \left[ a^2(t) \sin^2{\psi} + 
               b^2(t) \cos^2{\psi} \right] - c^2(t) \cos^2{\theta}   ,\\
g_{\1\3} & = & - c^2(t) \cos{\theta}   ,\\
g_{\2\3} & = & - \sin{\psi} \cos{\psi} \sin{\theta} 
           \left[ a^2(t) - b^2(t) \right]   .
\eea
The Ricci tensor components in the coordinate frame can be calculated using the
relations \cite{bkl} (cf. Eqs. (2.9)-(2.12))
\bea
R_{\alpha\beta} = e^{j}_{\alpha} e^{k}_{\beta} R_{jk}   ,\\
R_{\alpha}^{\beta} = e^{j}_{\alpha} e^{\beta}_{k} R_{j}^{k}   ,
\eea
where 
$R_{\alpha\beta}$ is the Ricci tensor in the coordinate frame while $R_{jk}$ is
the Ricci tensor in the orthonormal frame (correspondingly $\alpha, \beta =
\0, \1, \2, \3$ are the coordinate frame indices and $i, j = 0, 1, 2, 3$ are 
the orthonormal frame indices).
Thus, $1, 2, 3$ refer to $\sigma^1, \sigma^2, \sigma^3$, while $\bar{1}, 
\bar{2}, \bar{3}$ refer to $\psi, \theta, \varphi$ respectively.
According to (2.10)-(2.12), 
\bea
e^1_{\bar{2}} & = & a \cos{\psi}  \nonumber ,\\
e^1_{\bar{3}} & = & a \sin{\psi} \sin{\theta} \nonumber  ,\\
e^2_{\bar{2}} & = & b \sin{\psi}  ,\\
e^2_{\bar{3}} & = & - b \cos{\psi} \sin{\theta} \nonumber  ,\\
e^3_{\bar{1}} & = & c  \nonumber ,\\
e^3_{\bar{3}} & = & c \cos{\theta}   \nonumber,
\eea
and 
\bea
e^{\bar{1}}_1 & = & - \frac{1}{a} \sin{\psi} \cot{\theta}   ,\\
e^{\bar{1}}_2 & = &  \frac{1}{b} \cos{\psi} \cot{\theta}  \nonumber ,\\
e^{\bar{1}}_3 & = &  \frac{1}{c}  \nonumber ,\\
e^{\bar{2}}_1 & = &  \frac{1}{a} \cos{\psi}   ,\\
e^{\bar{2}}_2 & = &  \frac{1}{b} \sin{\psi}  \nonumber ,\\
e^{\bar{3}}_1 & = &  \frac{1}{a} \frac{\sin{\psi}}{\sin{\theta}}  \nonumber ,\\
e^{\bar{3}}_2 & = & - \frac{1}{b} \frac{\cos{\psi}}{\sin{\theta}}  \nonumber .
\eea

Then, the Ricci tensor components in the coordinate frame are given by 
\bea
R_{\bar{1}\bar{1}} & = & c^2 R_{33}   ,\\
R_{\bar{2}\bar{2}} & = & a^2 \cos^2{\psi} R_{11} + b^2 \sin^2{\psi} R_{22}   ,\\
R_{\bar{3}\bar{3}} & = & a^2 \sin^2{\psi} \sin^2{\theta} R_{11} + 
b^2 \cos^2{\psi} \sin^2{\theta} R_{22} + c^2 \cos^2{\theta} R_{33}   ,\\
R_{\bar{1}\bar{3}} & = & c^2 \cos{\theta} R_{33}   ,\\
R_{\bar{2}\bar{3}} & = & \sin{\psi} \cos{\psi} \sin{\theta} 
\left( a^2 R_{11} - b^2 R_{22} \right)   ,\\
R_{\bar{1}\bar{2}} & = & 0   ,
\eea
and
\bea
R_{\bar{1}}^{\bar{1}} & = & R_{3}^{3}   ,\\
R_{\bar{2}}^{\bar{2}} & = & \cos^2{\psi} R_{1}^{1} + \sin^2{\psi} R_{2}^{2}   ,\\
R_{\bar{3}}^{\bar{3}} & = & \sin^2{\psi} R_{1}^{1} + \cos^2{\psi} R_{2}^{2}   ,\\
R_{\bar{3}}^{\bar{1}} & = & \left[ - \sin^2{\psi} R_{1}^{1} - 
\cos^2{\psi} R_{2}^{2} + R_{3}^{3} \right] \cos{\theta}   ,\\
R_{\bar{3}}^{\bar{2}} & = & \sin{\psi} \cos{\psi} \sin{\theta} \left(
R_{1}^{1} - R_{2}^{2} \right)   ,\\
R_{\bar{2}}^{\bar{3}} & = & \frac{\sin{\psi} \cos{\psi}}{\sin{\theta}} 
\left(R_{1}^{1} - R_{2}^{2} \right)   ,\\
R_{\bar{2}}^{\bar{1}} & = & -\sin{\psi} \cos{\psi} \cot{\theta} \left(
R_{1}^{1} - R_{2}^{2} \right)   ,\\
R_{\bar{1}}^{\bar{3}} & = & 0   ,\\
R_{\bar{1}}^{\bar{2}} & = & 0   .
\eea

If two axes are the same ($a(t) = b(t)$) the metric (2.9) (or its components given by (A1)-(A6)) simplifies to  
\cite{brill,j2} 
\be
ds^2 = dt^2 - c^2 \left( d\psi + \cos{\theta}d\varphi \right)^2 - 
a^2 \left( d\theta^2 + \sin^2{\theta} d\varphi^2 \right)   .
\ee

The nonvanishing Christoffel symbols for the metric (A27) are 
\bea 
\Gamma^{\1}_{\0\1} = \frac{\dot{c}}{c} , \hspace{0.5cm} 
\Gamma^{\2}_{\0\2} = \frac{\dot{a}}{a} , \hspace{0.5cm} 
\Gamma^{\3}_{\0\3} = \frac{\dot{a}}{a}   , \nonumber \\
\Gamma^{\1}_{\0\3} = \left( \frac{\dot{c}}{c} - \frac{\dot{a}}{a} \right) 
\cos{\theta} , \hspace{0.5cm} \Gamma^{\0}_{\1\1} = \dot{c} c , \hspace{0.5cm} 
\Gamma^{\0}_{\2\2} = \dot{a} a   , \nonumber \\
\Gamma^{\0}_{\3\3} = \dot{c} c \cos^2{\theta} + \dot{a} a \sin^2{\theta} ,
\hspace{0.5cm} \Gamma^{\0}_{\1\3} = \dot{c} c \cos{\theta}   , \nonumber \\
\Gamma^{\1}_{\2\1} = \frac{1}{2} \frac{c^2}{a^2} \cot{\theta} , \hspace{0.5cm} 
\Gamma^{\1}_{\2\3} = \frac{1}{2\sin{\theta}} \left( \frac{c^2 - a^2}{a^2}
\cos^2{\theta} - 1 \right)   ,  \\
\Gamma^{\2}_{\3\1} = \frac{1}{2} \frac{c^2}{a^2} \sin{\theta} , \hspace{0.5cm} 
\Gamma^{\2}_{\3\3} = \sin{\theta} \cos{\theta} \frac{c^2 - a^2}{a^2}   , \nonumber
\\
\Gamma^{\3}_{\1\2} = - \frac{1}{2} \frac{c^2}{a^2} \frac{1}{\sin{\theta}} , \hspace{0.5cm} 
\Gamma^{\3}_{\2\3} = \cot{\theta} \left( 1 - \frac{1}{2} \frac{c^2}{a^2}
\right)  \nonumber .
\eea

The gradients of the dilaton calculated with respect to the metric (A27) are given by 
\bea
\nabla_{\0} \nabla^{\0} \phi & = & \ddot{\phi}   ,\\
\nabla_{\1} \nabla^{\1} \phi & = & \frac{\dot{c}}{c} \dot{\phi}   ,\\
\nabla_{\2} \nabla^{\2} \phi & = & \nabla_{\3} \nabla^{\3} = \frac{\dot{a}}{a} \dot{\phi}   ,\\
\nabla_{\3} \nabla^{\1} \phi & = & \dot{\phi} \left( \frac{\dot{c}}{c} - \frac{\dot{a}}{a} \right) 
\cos{\theta}   ,
\eea
and
\be
\nabla_{\0}\phi \nabla^{\0}\phi = \dot{\phi}^2   .
\ee

The nonzero components of the Ricci tensor in a coordinate frame for the metric (A27) can be obtained from (A12)-(A26) by putting $R_1^1 = R_2^2$ and using (2.17)-(2.20) for $a = b$ or directly using the Christoffel symbols. This gives 
\bea
R_{\bar{1}\bar{1}} = \ddot{c} c + 2 \dot{c} c \frac{\dot{a}}{a} + \frac{1}{2}
\frac{c^4}{a^4} , \hspace{0.5cm} R_{\bar{1}\bar{3}} = R_{\bar{1}\bar{1}} \cos{\theta}   , \nonumber \\
R_{\bar{2}\bar{2}} = \ddot{a} a + \dot{a}^2 + \dot{a} a \frac{\dot{c}}{c} + 1 - \frac{1}{2}
\frac{c^2}{a^2} , \hspace{0.5cm} R_{\bar{0}\bar{0}} = - \frac{\ddot{c}}{c} - 2
\frac{\ddot{a}}{a}   ,  \\
R_{\bar{3}\bar{3}} = R_{\bar{1}\bar{1}} \cos^2{\theta} + R_{\bar{2}\bar{2}}
\sin^2{\theta}   \nonumber ,
\eea
so 
\bea
- R_{\bar{0}}^{\bar{0}} & = & \frac{\ddot{c}}{c} + 2 \frac{\ddot{a}}{a}    ,\\
- R_{\bar{1}}^{\bar{1}} & = & \frac{\ddot{c}}{c} + 2 \frac{\dot{a}}{a} \frac{\dot{c}}{c} 
+ \frac{1}{2} \frac{c^2}{a^4}   ,\\
- R_{\bar{2}}^{\bar{2}} & = & - R_{\bar{3}}^{\bar{3}} = \frac{\ddot{a}}{a} +  \frac{\dot{a}^2}{a^2} + 
\frac{\dot{a}}{a} \frac{\dot{c}}{c} + \frac{1}{a^2} - 
\frac{1}{2} \frac{c^2}{a^4}   ,\\
R_{\bar{3}}^{\bar{1}} & = & \left( \frac{\ddot{a}}{a} - \frac{\ddot{c}}{c} - 
\frac{\dot{a}}{a} \frac{\dot{c}}{c} + \frac{\dot{a}^2}{a^2} + \frac{1}{a^2} - 
\frac{c^2}{a^4} \right) \cos{\theta}   ,
\eea
and the Ricci scalar reads
\be
R = - 2 \frac{\ddot{c}}{c} - 4 \frac{\ddot{a}}{a} - 
4 \frac{\dot{a}}{a} \frac{\dot{c}}{c} - 2 \frac{\dot{a}^2}{a^2} - 
2 \frac{1}{a^2} + \frac{1}{2} \frac{c^2}{a^4} 
\ee

Note that in the coordinate frame $ R_{\bar{2}}^{\bar{2}} = 
 R_{\bar{3}}^{\bar{3}}$ while in the orthonormal frame (cf. Eq.(2.18)-(2.19)) 
$R_1^1 = R_2^2$. This is reasonable, since the metric tensor is given by (A27),
for the former case, and by (2.9) with $a = b$ for the latter case, where the
indices refer to the orthonormal basis rather than to the chosen coordinates.

Now we convert our notation in terms 
of the three-index torsion field $H$ to the notation given in \cite{ed} using 
the pseudoscalar torsion field $h$. Following \cite{ed} we define ($\alpha, \beta, \mu, \nu = \0, \1, \2, \2$) 
\be
H^{\mu\nu\alpha} = e^{\phi} \epsilon^{\mu\nu\alpha\beta} h_{,\beta}   .
\ee
The equation of motion for $h$-field obtained via integrability conditions is then 
\be
\nabla^{\mu}\nabla_{\mu} h + \nabla^{\mu}\phi \nabla_{\mu} h = 0   .
\ee
One can easily see that for the time-independent antisymmetric tensor potential $B_{\mu\nu} = 
B_{\mu\nu}(x)$ the $h$ field can only depend on time and the equation (A41) reads as 
\be
\ddot{h} + \left( \frac{\dot{a}}{a} + \frac{\dot{b}}{b} + \frac{\dot{c}}{c} 
 \right) \dot{h} + 
\dot{\phi} \dot{h} = 0   ,
\ee
which integrates to give 
\be
\dot{h} = - A \frac{e^{- \phi}}{abc}   ,
\ee
so from (A40) we have 
\be
H^{\1\2\3} = - \frac{A}{a^2b^2c^2 sin{\theta}}  ,
\ee
or 
\be
H_{\1\2\3} = A sin{\theta}   ,
\ee
as required by (2.25) and $H^2 = - 6A^2/a^2b^2c^2$. With the $H$ field chosen as 
above the equation of motion 
(2.3) is easily fulfilled. There is also a trivial solution of (A42), 
$\dot{h} = 0$, but it corresponds to a constant torsion field.
For the time-independent pseudoscalar axion field $h = h(x)$ (time-dependent antisymmetric tensor potential $B_{\mu\nu} = B_{\mu\nu}(t)$) the equation of motion (A41) reads
\be
\partial^{\mu}\partial_{\mu} h + \Gamma^{\mu}_{\rho \mu} \partial^{\rho}h = 0   ,
\ee
which for the metric (2.9) reads as 
\be
g^{\1\1} \partial^{2}_{\1} h + g^{\2\2} \partial^{2}_{\2} h 
+ g^{\3\3} \partial^{2}_{\3} h + g^{\1\3} \partial_{\1} \partial_{\3} h 
+ g^{\2\2} \cot{\theta} \partial_{\2}h = 0   .
\ee
For simplicity, let us introduce 
\bea
(H^{\nu}_{\mu})^2 & = & H_{\mu\alpha\beta} H^{\nu\alpha\beta} = 
- 2 e^{2\phi} \left( \delta_{\mu}^{\nu} g^{\rho\varepsilon} 
- \delta_{\mu}^{\rho} g^{\nu\varepsilon} \right) \partial_{\varepsilon}h 
\partial_{\rho}h   ,\\
H^{2} & = & H_{\mu\alpha\beta} H^{\mu\alpha\beta}   .
\eea
The non-zero components of these quantities (the energy-momentum tensor) which 
are used in the field equations (2.1)-(2.3) for the metric components
(A27) are given by 
\bea
(H^{\0}_{\0})^2 & = & - 2 e^{2 \phi} \left( g^{\1\1} \partial_{\1} h \partial_{\1} h
+ 2 g^{\1\3} \partial_{\1} h \partial_{\3} h + g^{\2\2} \partial_{\2} h \partial_{\2} h  
+ g^{\3\3} \partial_{\3} h \partial_{\3} h \right)    ,\\
(H^{\1}_{\1})^2 & = & - 2 e^{2 \phi} \left( g^{\1\3} \partial_{\1} h \partial_{\3} h 
+ g^{\2\2} \partial_{\2} h \partial_{\2} h  
+ g^{\3\3} \partial_{\3} h \partial_{\3} h \right)    ,\\
(H^{\2}_{\2})^2 & = & - 2 e^{2 \phi} \left( g^{\1\1} \partial_{\1} h \partial_{\1} h
+ 2 g^{\1\3} \partial_{\1} h \partial_{\3} h + g^{\3\3} \partial_{\3} h \partial_{\3} h 
\right)    ,\\
(H^{\3}_{\3})^2 & = & - 2 e^{2 \phi} \left( g^{\1\1} \partial_{\1} h \partial_{\1} h
+ g^{\1\3} \partial_{\1} h \partial_{\3} h + g^{\2\2} \partial_{\2} h \partial_{\2} h  
\right)    ,\\
(H^{\1}_{\3})^2 & = & 2 e^{2 \phi} \left( g^{\1\1} \partial_{\1} h \partial_{\3} h
+ g^{\1\3} \partial_{\3} h \partial_{\3} h \right)    ,
\eea
and
\be
H^2 = - 6 e^{2 \phi} \left( g^{\1\1} \partial_{\1} h \partial_{\1} h
+ 2 g^{\1\3} \partial_{\1} h \partial_{\3} h + g^{\2\2} \partial_{\2} h \partial_{\2} h  
+ g^{\3\3} \partial_{\3} h \partial_{\3} h \right)    ,
\ee
With the choice 
\be
\partial_{\3} h = E \sin{\theta} \neq 0 \hspace{2.cm} \partial_{\1} h 
= \partial_{\2} h = 0   ,
\ee
the field equations remain homogeneous and 
\bea
(H_{\0}^{\0})^2 & = & (H_{\1}^{\1})^2 = (H_{\2}^{\2})^2  = 2 \frac{E^2 e^{2\phi}}{a^2}
,\\
(H_{\3}^{\3})^2 & = & 0    ,\\
(H_{\3}^{\1})^2 & = & 2 \frac{E^2 e^{2\phi}}{a^2} \cos{\theta}   ,\\
H^2 & = & 6 \frac{E^2 e^{2\phi}}{a^2}
\eea
and the equation of motion (A47) becomes 
\be 
g^{\3\3} \partial_{\varphi}(E\sin{\theta}) = 0   ,
\ee
and it is satisfied. 
Finally, from (A40), we have
\bea
H^{\0\1\2} = \frac{E e^{\phi}}{a^2 c}  ,\\
H_{\0\1\2} = E c e^{\phi}   ,
\eea
and $H^2 = 6E e^{2\phi}/a^2$. 
This obeys the axion equation of motion (2.3) but is in contradiction with the
axisymmetry condition for the Ricci components (A37) since $R_2^2 = R_3^3$ 
there while $(H_{\3}^{\3})^2 = 0$ and $(H_{\2}^{\2})^2 \neq 0$ here.

A possible ansatz which would fulfil the axisymmetry condition would be 
\be
\partial_{\1} h = \frac{B a(t) \sin{\theta}}{\sqrt{a(t)^2 \sin^2{\theta} + 
c(t)^2 \cos^2{\theta}}}   ,
\ee
but it leads to both time and space dependences of the pseudoscalar axion
field, $h = h(t,\psi,\theta)$, and does not fulfil the equation of motion (A41) 
(nor (A47) which is obtained for $h = h(x)$).

One could also try to add
\be
\partial_{\2} h = D = {\rm const.}   ,
\ee
to the nonzero component (A56), but again this does not fulfil the 
equation (A47). 

The final conclusion is that one is not able to impose the axion field even in 
axisymmetric BIX models despite the fact that there is a distinguished direction
in the model (which is different from electromagnetic field case - see
\cite{brill}). The reason seems to be that even in the axisymmetric 
case there is still $SO(3)$ symmetry group present and we are only adding an  additional
symmetry $SO(2)$ which does not cancel the former one, giving the total symmetry
$SO(3) \bigotimes SO(2)$ rather than just $SO(2)$.

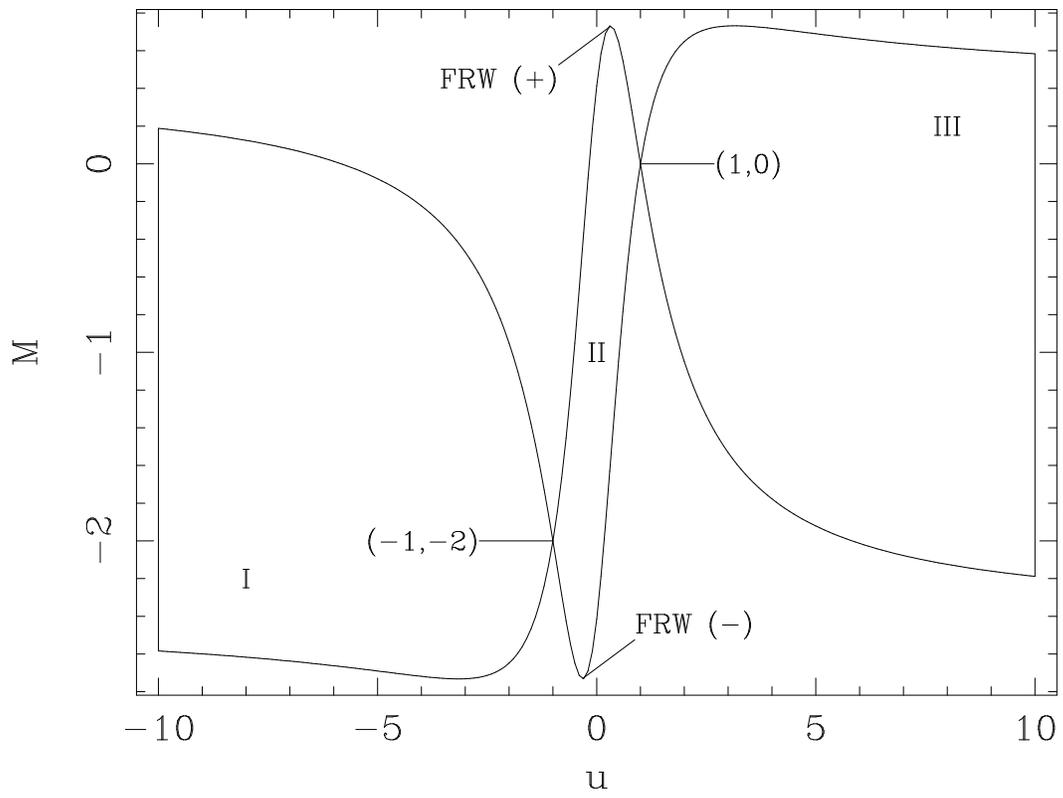
\begin{figure}
\caption[]
{The regions I, II, III of permissible values of the parameters $u$ and $M$ for the parametrization (3.60) of the Kasner indices $p_1, p_2$ and $p_3$. We restricted ourselves to $ -10 \leq u \leq 10$ although regions I and III extend to infinity. Special isotropic FRW (+) and FRW (-) points are given for $(\sqrt{3} - \sqrt{2}, - 1 + \sqrt{3})$ and $(- \sqrt{3} + \sqrt{2}, - 1 - \sqrt{3})$ respectively.}
\label{fig1}
\end{figure}

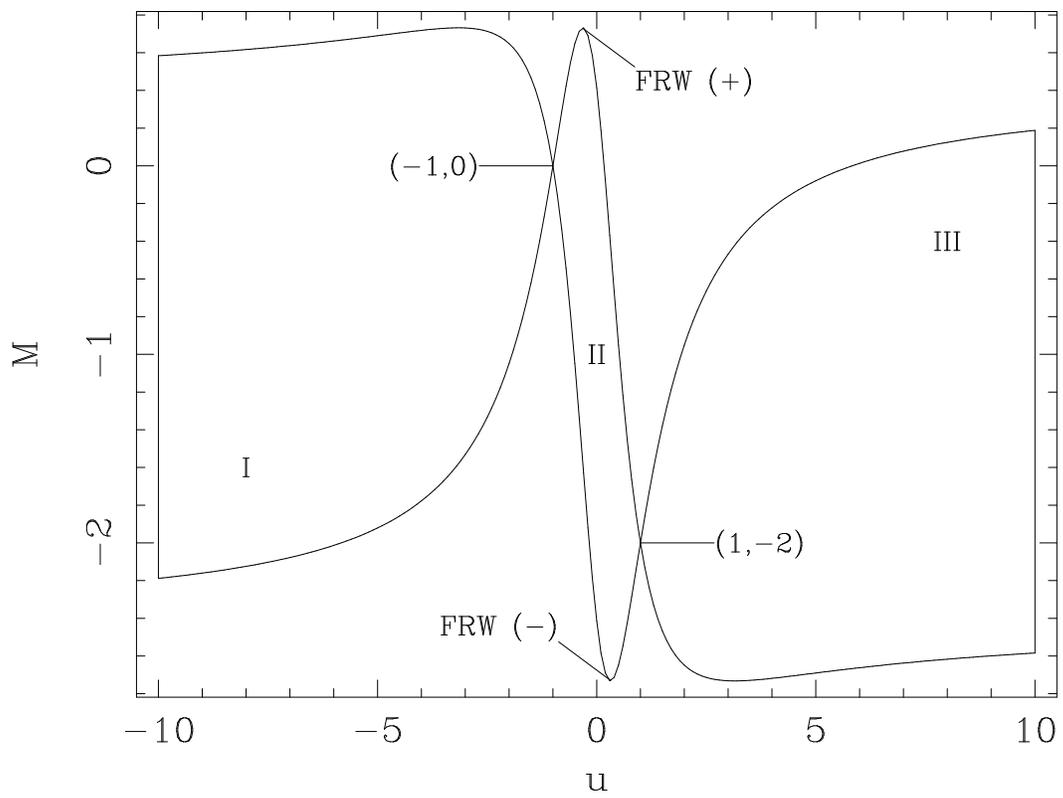
\begin{figure}
\caption[]
{The regions I, II, III of permissible values of the parameters $u$ and $M$ for the parametrization (3.62) of the Kasner indices $p_1, p_2$ and $p_3$. We restricted ourselves to $ -10 \leq u \leq 10$ although regions I and III extend to infinity. Special isotropic FRW (+) and FRW (-) points are given for $(-\sqrt{3} + \sqrt{2}, - 1 + \sqrt{3})$ and $(\sqrt{3} - \sqrt{2}, - 1 - \sqrt{3})$ respectively.}
\label{fig2}
\end{figure}

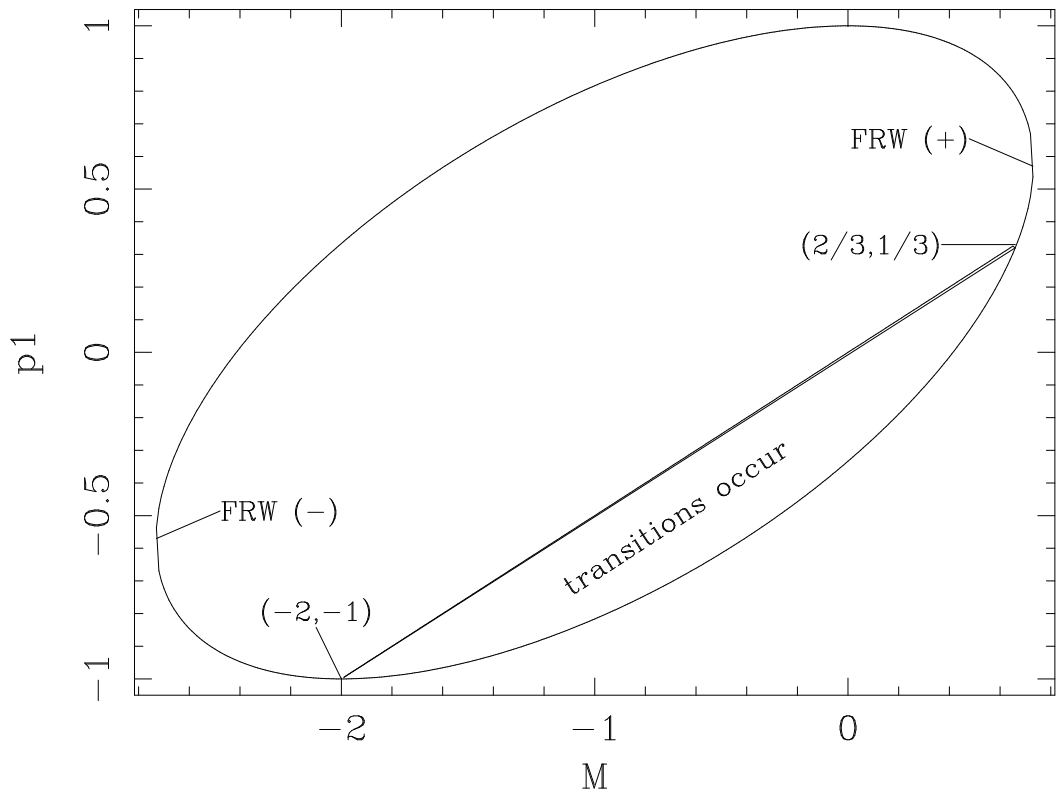
\begin{figure}
\caption[]
{The plot of the conditions (3.96)-(3.97) for the transitions from one Kasner epoch to another to begin in terms of the first and the "fourth" Kasner indices  $p_1$ and $M$. Obviously, isotropic FRW (+) and FRW (-) points $(-1 + \sqrt{3}, 1/\sqrt{3})$ and $(-1 - \sqrt{3}, -1/\sqrt{3})$ respectively, are excluded. The transitions occur below the line $p = M/2$. From the picture one can easily see  that in vacuum general relativity case ($M = 0$) the transitions are possible whenever $ - 1/3 \leq p_1 \leq 0$. This is in agreement with the standard calculations. However, in the dual case $(M = - 2)$ transitions do not occur at all.} 
\label{fig3}
\end{figure} 

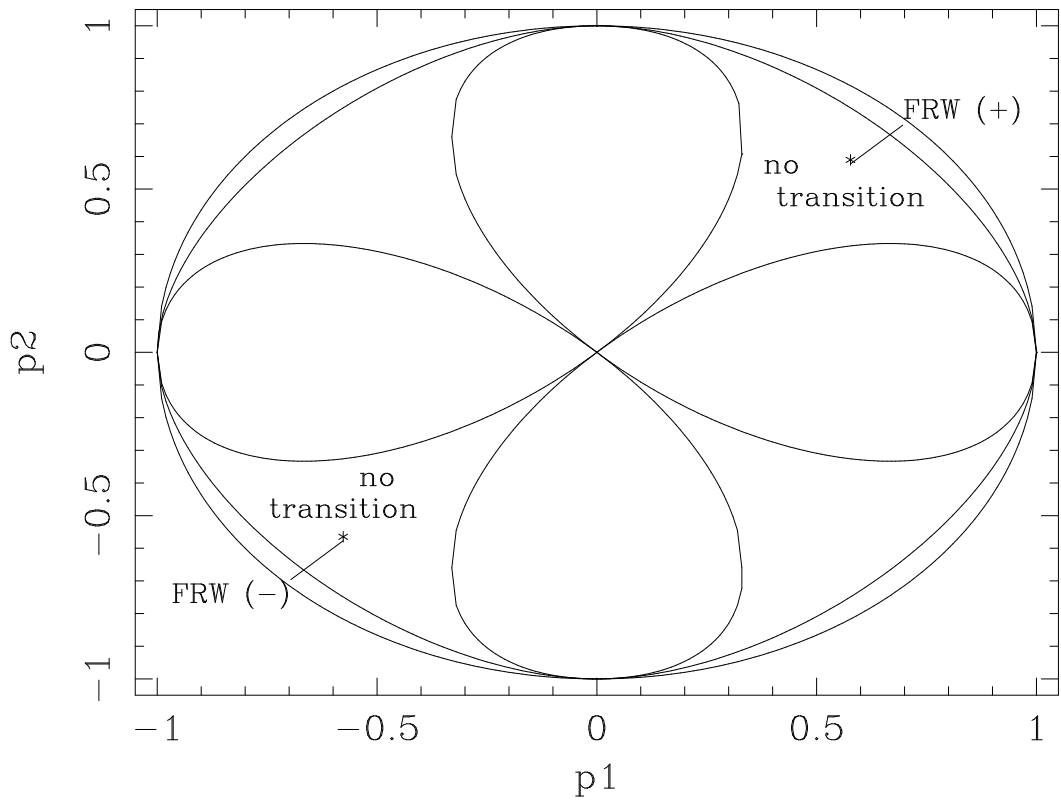
\begin{figure}
\caption[]
{The plot of the conditions (3.101)-(3.103) for the transitions from one Kasner epoch to another to begin, in terms of the two Kasner indices $p_1$ and $p_2$. As in Fig. 3 the isotropic FRW (+) and FRW (-) points, $(1/\sqrt{3}, 1/\sqrt{3})$ and $(-1/\sqrt{3}, -1/\sqrt{3})$ respectively, together with the two neighbouring regions, are excluded.}
\label{fig4}
\end{figure}

\pagebreak
Figure 1: \hfill

\vskip2cm
\centerline{\psfig{figure=fig1.ps,height=14cm} \hskip 2cm}

\pagebreak
Figure 2: \hfill

\vskip2cm
\centerline{\psfig{figure=fig2.ps,height=14cm} \hskip 2cm}

\pagebreak
Figure 3: \hfill

\vskip2cm
\centerline{\psfig{figure=fig3.ps,height=14cm} \hskip 2cm}

\pagebreak
Figure 4: \hfill

\vskip2cm
\centerline{\psfig{figure=fig4.ps,height=14cm} \hskip 2cm}


\begin{thebibliography}{99}
\bibitem{fradcall} E.S. Fradkin and A.A. Tseytlin, Nucl. Phys. B{\bf 261}, 
               1 (1985);\\
               C.G. Callan, D. Friedan, E.J. Martinec and M.J. Perry, Nucl.
               Phys B{\bf 262}, 593 (1985).
\bibitem{strmod}  M. Mueller, Nucl. Phys. B{\bf 37}, 37 (1990);\\
               G. Veneziano, Phys. Lett. B{\bf 265}, 287 (1991);\\
               M. Gasperini and G. Veneziano, Mod. Phys. Lett. {\bf 39},
               3701 (1993);\\
               M. Gasperini and G. Veneziano, Astro. Phys. {\bf 1}, 317
               (1993);\\
               E.J. Copeland, A. Lahiri and D. Wands, Physical Review 
               D{\bf 50}, 4868 (1994);\\
               M. Gasperini and R. Ricci, Class. Quantum Grav. {\bf 12}, 
               677 (1995);\\
               A. Saaryan, Astrophysics {\bf 38}, 164 (1995);\\  
               M. Gasperini and G. Veneziano, Phys. Rev. D{\bf 50},
               2519 (1994).                                         
\bibitem{ed} E.J. Copeland, A. Lahiri and D. Wands, Physical Review  
             D{\bf 51}, 1569 (1995).
\bibitem{homog} J.D. Barrow and K.E. Kunze, Phys. Rev. D{\bf 55}, 623 (1997);\\ 
                J.D. Barrow and M.P. D\c{a}browski, Phys. Rev. D{\bf 55}, 630 
                (1997).
\bibitem{bk} N.A. Batakis and A. Kehagias, Nucl. Phys. B{\bf 449}, 
             248 (1995), J.P. Mim\'oso and D. Wands, Phys. Rev. D{\bf 52}, 5612 
             (1995).
\bibitem{bkl} V.A. Belinskii, E.M. Lifshitz and I.M. Khalatnikov, Sov. Phys.
         Uspekhi {\bf 102}, 745 (1971);\\
         L.D. Landau and E.M. Lifshitz, {\it The Classical Theory of Fields},
         Oxford (1975);\\
         J.D. Barrow, Phys. Rev. Lett. {\bf 46}, 963 (1981);\\
         J.D. Barrow, Phys. Rep. {\bf 85}, 1 (1982);\\
         D. Chernoff and J.D. Barrow, Phys. Rev. Lett. {\bf 50}, 134 (1983);\\
         J.D. Barrow, in {\it Classical General Relativity}, ed. W. Bonnor, J.  
         Islam, and M.A.H. MacCallum, CUP, Cambridge (1984), p.25.
\bibitem{belkhal} V.A. Belinskii and I.M. Khalatnikov, Sov. Phys. JETP {\bf 36},
        591 (1973).
\bibitem{veldom} G. Veneziano, Phys. Lett. B{\bf 406}, 297 (1997);\\
               A. Buonanno, K.A. Meissner, C. Ungarelli and G. Veneziano 
              `Classical Inhomogeneities in String Cosmology', e-Print:
               hep-th/9706221;\\ 
               A. Feinstein, R. Lazkoz and M.A. V\'azquez-Mozo `Closed 
               Inhomogeneous String Cosmologies', e-Print: hep-th/9704173.
\bibitem{barkun} J.D. Barrow and K.E. Kunze, Phys. Rev. D{\bf 56}, 741 (1997).
\bibitem{linde} B.A. Campbell, A. Linde and K.A. Olive, Nucl. Phys. B{\bf
         355}, 146 (1991); Eq.(7). 
\bibitem{brill} D.R. Brill, Phys. Rev. {\bf 133}, B845 (1964).
\bibitem{j2} J.D. Barrow, Nature {\bf 272}, 211 (1978).
\bibitem{j1} J.D. Barrow, Nucl. Phys. B{\bf 296}, 697 (1988).
\bibitem{dave1} R. Easther, K. Maeda and D. Wands, Phys. Rev. {\bf D53}
         (1996) 4247.
\bibitem{gv} M. Gasperini and G. Veneziano, {\it Gen. Rel. Grav.} {\bf 28}, 
         1301 (1996).
\bibitem{wainwright} J. Wainwright and L. Hsu, Class. Quantum Grav., {\bf 6},
        1409 (1989);\\
         J. Wainwright, {\it Dynamical Systems in Cosmology}, Ed. J. Wainwright 
         and G. Ellis (CUP, Cambridge, 1997), p.132.
\bibitem{ryan} M.P. Ryan, L.C. Shepley, {\it Homogeneous Relativistic 
         Cosmologies} (Princeton University Press, Princeton, 1975).
\bibitem{pade} J.S.R. Chrisholm and J. McEwan, Proc. Roy. Soc. {\bf A336}, 421 
        (1974);\\
         G.A. Baker, {\it Essentials of Pad\'e Approximants} (Academic 
         Press, New York, 1975).
\bibitem{halp} P. Halpern, Phys. Rev. D{\bf 33}, 354 (1986).
\bibitem{john85} J.D. Barrow and J. Stein-Schabes, Phys. Rev. D{\bf32}, 1595
        (1985).
\bibitem{ishihara} H. Ishihara, Prog. Theor. Phys. {\bf 74}, 354 (1986).
\bibitem{hosoya} T. Furosawa and A. Hosoya, Prog. Theor. Phys. {\bf 73}, 467  
        (1985).
\bibitem{james} J.E. Lidsey, 'On the Cosmology and Symmetry of Dilaton-Axion 
        Gravity', e-Print gr-qc/9609063.
\bibitem{bframe} E.J. Copeland, R. Easther and D. Wands, 'Vacuum Fluctuations
        in Axion-Dilaton Cosmologies', e-Print: hep-th/9701082.
\bibitem{andrew} A.R. Liddle, R.G. Moorhouse and A.B. Henriques, Nucl. Phys.
        B{\bf 311}, 719 (1988).
\bibitem{Mtheory} D.V. Gal'tsov and M.S. Volkov, Phys. Lett. B{\bf 256}, 17 
        (1991);\\
         I. Ya. Aref'eva, P.B. Medvedev, O.A. Rytchkov and I.V.
         Volovich, `Chaos in M(atrix) Theory', e-Print:hep-th/9710032;\\
         J.D. Barrow and J. Levin, `Chaos in the Einstein-Yang-Mills Equations'
         to appear in Phys. Rev. Lett.


\end{thebibliography}
\end{document}